\documentclass[12pt]{article} 
\pdfoutput=1

\usepackage{graphicx} 
\usepackage{slashed}
\usepackage[margin=.9in]{geometry}
\usepackage{amsmath,amssymb,amsfonts,wasysym}
\usepackage[normal,font=small,labelfont=bf,labelsep=period]{caption}
\usepackage[compress]{cite}
\usepackage{xcolor}
\usepackage[linktoc=page,colorlinks=true,linkcolor=blue,citecolor=darkred,urlcolor=blue]{hyperref}
\usepackage[margin=.9in]{geometry}
\definecolor{darkblue}{rgb}{0,0.2,0.6}
\definecolor{darkred}{rgb}{0.9,0.2,0.2}

\newcommand{\ga}{SM$\times$SM$'$\,}
\newcommand{\nn}{\nonumber}

\renewcommand{\div}{-}
\newcommand{\trans}{T}
\newcommand{\order}{\mathcal{O}}

\title{\vspace{-2cm}
\begin{flushright}
{\small EFI-15-2}\vspace{1cm}
\end{flushright}
\noindent
{\Large \bf The Twin Higgs mechanism and Composite Higgs}}

\author{Matthew Low$^a$, Andrea Tesi$^b$, Lian-Tao Wang$^a$}
\date{\it \small $^a$Department of Physics, Enrico Fermi Institute, and Kavli Institute for\\
Cosmological Physics, University of Chicago, Chicago, IL 60637\\
\small $^b$Department of Physics, Enrico Fermi Institute, University of Chicago, Chicago, IL 60637}

\begin{document}
\begin{titlepage}

\maketitle
\thispagestyle{empty}

\begin{abstract}
\noindent
We combine the Twin Higgs mechanism with the paradigm of Composite Higgs models. In this class of models the Higgs is a pseudo-Nambu-Goldstone boson from a strongly coupled sector near the TeV scale, and it is additionally protected by a discrete symmetry due to the twin mechanism. We discuss the model building issues associated with this setup and quantify the tuning needed to achieve the correct electroweak vacuum and the Higgs mass. In contrast to standard Composite Higgs models, the lightest resonance associated with the top sector is the uncolored mirror top, while the colored top partners can be made parameterically heavier without extra tuning.  In some cases, the vector resonances are predicted to lie in the multi-TeV range.  We present models where the resonances -- both fermions and vectors -- being heavier alleviates the pressure on naturalness coming from direct searches demonstrating that theories with low tuning may survive constraints from the Large Hadron Collider.
\end{abstract}

\vfill
\noindent\line(1,0){188}\\\medskip
\footnotesize{E-mail: \texttt{\href{mailto:mattlow@uchicago.edu}{mattlow@uchicago.edu}, \href{mailtoatesi@uchicago.edu}{atesi@uchicago.edu}, \href{mailto:liantaow@uchicago.edu}{liantaow@uchicago.edu}}}

\end{titlepage}

\newpage
\tableofcontents
\section{\label{sec:intro}Introduction}

There are several possibilities to naturally stabilize the electroweak scale and the Higgs mass against large UV corrections. However, after the discovery of a light Standard Model (SM)-like Higgs boson and increasing limits on new particles from the first run of the LHC, it is hard to find models less tuned than $\sim 5\div 10 \%$. The challenges faced by fully natural models are becoming several, one of them in particular is the lack of any sign of new colored particles around the weak scale.  This is particularly true for the case of Natural SUSY~\cite{Papucci:2011wy,Hall:2011aa}, where light stops are generally required. 

On the other hand, in strongly coupled scenarios such as Composite Higgs (CH), where the Higgs is a pseudo-Nambu-Goldstone boson (pNGB) of a given coset $G/H$~\cite{Dugan:1984hq}, new colored vector-like quarks are expected to lie within a few hundreds of GeV of the Higgs. More specifically they are expected to be close to $f$, where $f$ is the Goldstone scale.  This is a direct consequence of the partial compositeness mechanism implemented in these models to solve the flavor problem~\cite{Kaplan:1991dc, Contino:2006nn}.  In this framework, the SM quarks are an admixture of elementary quarks and composite fermions, which we will write generically as $\Psi$, with the same SM gauge quantum numbers.  In this case, the Higgs mass is~\cite{Agashe:2004rs,Contino:2006qr, Contino:2010rs} 
\begin{equation}\label{predictionsCH}
m_h^2 \simeq \frac{N_c y_t^2 v^2}{2\pi^2}\frac{m_\Psi^2}{f^2}.
\end{equation}
From these estimates, as well as from several other checks~\cite{Matsedonskyi:2012ym,Redi:2012ha,Marzocca:2012zn,Pomarol:2012qf}, the prediction of natural CH models is the presence of light fermionic resonances, \textit{i.e.} with a mass $m_\Psi \sim f$.  This prediction also applies to electroweak composite vector resonances, but their masses can be larger than $m_\Psi$ due to the smallness of the weak gauge coupling, $g$, compared to $y_t$.  While this simple scenario points to straightforward and testable LHC signals, it is useful to assess the robustness of such a connection. There are, however, composite Higgs scenarios where one can deviate from this conclusion, {\it i.e.} models where $m_h=125$ GeV without the need for light top partners, but those models are severely fine tuned~\cite{Panico:2012uw}. 

An interesting possibility to disentangle, without additional tuning, the strong connection between a light Higgs and light colored top partners, while keeping the scale $f$ as small as possible,\footnote{An orthogonal possibility is offered by the Little Higgs mechanism, where the model building allows for a larger separation between $f$ and $v$ without large tuning~\cite{Schmaltz:2005ky} (see references therein), but we will not discuss them here.  See also~\cite{Bellazzini:2014yua} for a comparison between Little Higgs and Composite Higgs.} can be offered by the Twin Higgs (TH) mechanism~\cite{Chacko:2005pe}, see also \cite{Chacko:2005un}. As far as the SM alone is considered, the TH mechanism consists of simply mirroring the SM lagrangian, via a $Z_2$-symmetry, resulting in the SM and its copy, SM$'$~\cite{Barbieri:2005ri}. The scalar potential has an accidental U(4)/U(3) symmetry, and radiative corrections do not break it at the level of the quadratic action thanks to the $Z_2$ symmetry~\cite{Chacko:2005pe}.  The embedding of this mechanism in calculable models, like CH models (see~\cite{Falkowski:2006qq,Chang:2006ra,Craig:2013fga} for the supersymmetric case), serves as a test as to how well the twin mechanism works to realize the weak scale with low tuning.\footnote{See also orbifold models~\cite{Craig:2014aea,Craig:2014roa}.}

\begin{figure}[h!]
\begin{center}
\includegraphics{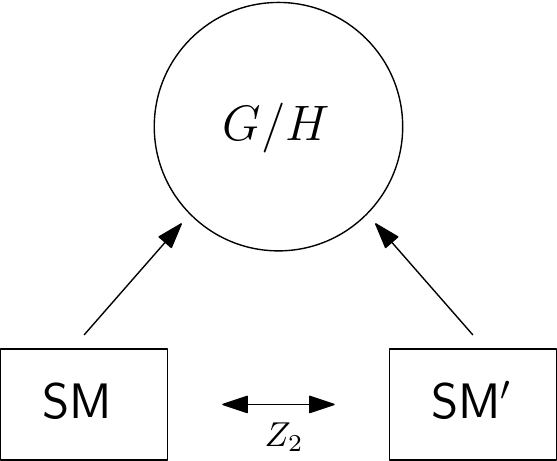}
\caption{\label{fig:picture} Pictorial representation of the dynamics of a Composite Higgs model protected by the Twin Higgs mechanism. In the model under consideration, $G/H=$~SO(8)/SO(7). This global symmetry is explicitly broken by the interactions with the two external copies of the SM (exchangeable under a $Z_2$ symmetry).}
\end{center}
\end{figure}

As we will show in this paper, CH models augmented with the TH mechanism provide a correct Higgs mass with minimal tuning and without any light colored top partner (see also \cite{Burdman:2014zta}).\footnote{Or conversely, TH models embedded inside the CH framework offer a more UV friendly setup.} In this scenario the lightest top partner is an uncolored mirror top from the mirror sector, with mass $\sim y_t f$, which replaces $\Psi$ in the prediction of eq.~\eqref{predictionsCH}. At a practical level, the simplest example of a Composite Twin Higgs (CTH) -- just a larger class of CH models -- relies on the global symmetry breaking SO(8)/SO(7). The global symmetry is explicitly broken by couplings to \ga.  As in standard CH, the explicit breaking induces radiative electroweak symmetry breaking (EWSB). The novelty here is that, roughly speaking, the overall scale of the potential is suppressed thanks to the protection of the additional $Z_2$ symmetry. This crucially depends on the coupling to a mirror elementary sector (both to gauge fields and fermions), as shown in figure~\ref{fig:picture}. However, in the limit of an exact $Z_2$ symmetry, we expect $v = f$. Due to the strong constraints on $f$, having $v \ll f$ is necessary. Hence, $Z_2$ needs to be broken to provide realistic realizations of CTH models.  

\begin{figure}[h!]
\begin{center}
\includegraphics[width=.5\textwidth]{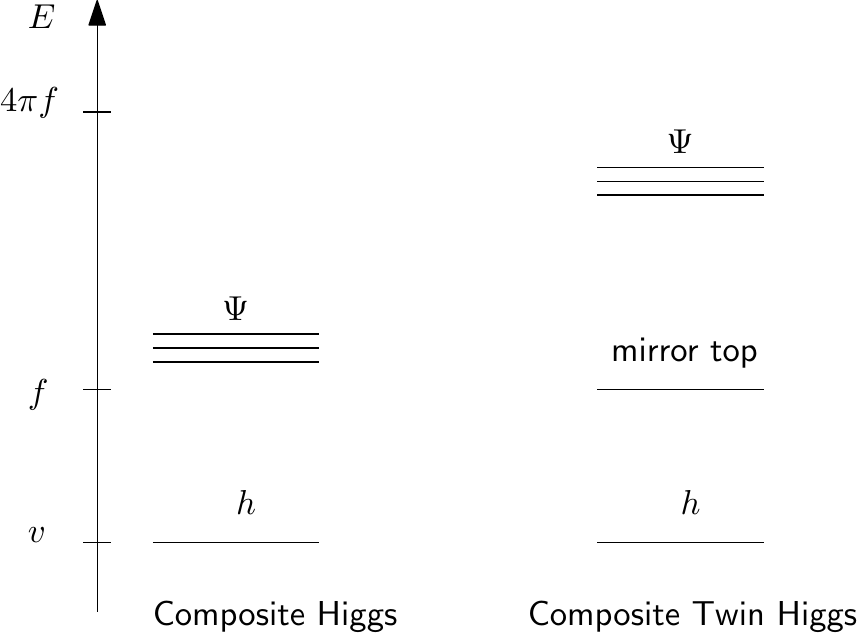}
\caption{\label{spectrum}\small Comparison between the spectra of minimally tuned Composite Higgs and Composite Twin Higgs models. }
\end{center}
\end{figure}

Depending on the actual breaking of $Z_2$ the Higgs potential can be relatively insensitive to the mass of the composite fermions, while being set by the scale of the uncolored mirror top.  If this is the case we can achieve the correct Higgs mass without colored top partners and trigger EWSB with minimal tuning.  A key result of this paper is to point out a class of models with this property.  A schematic drawing of the resulting spectrum, as well as a comparison with usual composite Higgs models, for minimal tuning $f^2/v^2$, is shown in figure~\ref{spectrum}. We provide estimates for the Higgs mass and tuning, as well several examples where we will be able to explicitly compute the Higgs potential. In order to do this we rely on purely four dimensional models (see~\cite{Geller:2014kta} for a holographic realization). To emphasize what is truly related to the TH mechanism, as opposed to standard CH models, we will consider several fermionic representations for the elementary quarks in order to show their effect on the Higgs potential.

The rest of the paper is organized as follows. In section~\ref{sec:model} we define the framework, discussing the coset structure, the gauging of \ga, fermion representations, and the form of the Higgs potential. A general parameterization of the Higgs potential as well as possible mechanisms of $Z_2$-breaking will be shown in section \ref{sec:ewsb}, after a brief introduction of the basic points in section \ref{sec:tuning}. Two concrete examples are provided in section \ref{sec:concrete}. We discuss the main phenomenology in section \ref{sec:pheno} and we conclude in section \ref{sec:conclusions}. We refer to appendix \ref{sec:app} for technical details.

\section{\label{sec:model}The SO(8)/SO(7) model}

The global symmetry breaking pattern of the simplest CTH model is SO(8)/SO(7).\footnote{The minimal coset used in linear realizations is U(4)/U(3), which delivers the same number of pNGBs, but does not contain a residual custodial symmetry.  Moreover, with U(4)/U(3) in the non-linear case the twin mechanism is not realized in the gauge sector (as first observed in~\cite{Barbieri:2015lqa}).  Groups larger than SO(8) could be fine as well, but there one expects the presence of extra physical pNGBs in addition to the Higgs.  An earlier work~\cite{Batra:2008jy} recognized the usefulness of SO(8)/SO(7) to prevent a large custodial breaking in composite models implementing the twin mechanism.  This model differs from ours in that it is based on left-right symmetry where the top partners are still colored under SM color.} The subgroup SO(7) allows for an unbroken SO(4) custodial group. The 7 pNGBs are encoded in the field $U$ obtained by the exponentiation of the fluctuations associated with the broken generators,
\begin{equation}\label{U}
U= \exp i \frac{\Pi}{f}.
\end{equation}
Given the basis of generators chosen in appendix \ref{sec:app}, the pNGB matrix $\Pi$ containing the 7 goldstones can be written as
\begin{equation}
\Pi= \sqrt{2} \pi^{\hat{a}} T^{\hat{a}}, \quad\quad\quad \hat{a}=1,\dots,7,
\end{equation}
where $T^{\hat{a}}$ are the broken generators, defined in appendix \ref{sec:app}, and $\pi^{\hat{a}}$ are the goldstone fields in the \textbf{7} of SO(7).  The transformation under SO(8)
\begin{equation}
U\to g\cdot U \cdot h(g,\Pi)^\trans, \nn
\end{equation}
lets us write the two indices of $U$ as $U_i^{\,\bar{j}}$ (we follow the notation of~\cite{DeSimone:2012fs}). The index $i$ is linear under $G$, while $\bar{j}=\{J, 8\}$ is non-linear under $G$ but split into a \textbf{7} of SO(7) (index $J$) and a singlet. Later we will make use of $U_i^{\,J}$ and $\Sigma_i \equiv U_i^{\,8}$,
\begin{equation}\label{sigma-full}
\Sigma^T =  \frac{\sin \frac{\pi}{f}}{\pi}(\pi^{\hat{1}},\pi^{\hat{2}},\pi^{\hat{3}},\pi^{\hat{4}},\pi^{\hat{5}},\pi^{\hat{6}},\pi^{\hat{7}},\pi \cot \tfrac{\pi}{f}), \quad\quad \pi \equiv \sqrt{\pi^{\hat{a}}\pi^{\hat{a}}}.
\end{equation} 
However, in CTH these are not the only global symmetries of the composite sector. Indeed, to realize the TH mechanism we have to include at the level of the composite sector a mirror copy of QCD, which amounts to having an unbroken $\mathrm{SU(3)}_c\times \mathrm{SU(3)}_c' \times Z_2$\footnote{Note that this is in contrast to orbifold-based models in which QCD and mirror QCD descend from an SU(6) group or larger~\cite{Craig:2014roa,Geller:2014kta}.}. Formally this means that the global symmetry is actually\footnote{There must be an unbroken U(1)$_X$ to obtain the correct hypercharge for each SM fermion.  It is defined by $Y=T^3_R+ X$ separately for both the sectors, where $T^3_R$ comes from each SU(2)$_R$ of each SO(4). The charges are $2/3$ and $-1/3$ for up-type and down-type quarks, respectively.  We similarly omit any discussion of the leptons as their effect on the Higgs potential is typically negligible.}
\begin{equation}
\frac{G}{H}=\frac{\mathrm{SU(3)}_c\times \mathrm{SU(3)}_c' \times Z_2 \times \mathrm{SO(8)}}{\mathrm{SU(3)}_c\times \mathrm{SU(3)}_c' \times Z_2 \times \mathrm{SO(7)}}.
\end{equation}
This guarantees that we can partially gauge the global symmetry $G$ by two identical copies of the SM, \ga (including QCD and its mirror copy). The gauging of \ga proceeds in the following way (see appendix \ref{sec:app} for details): we identify the two SO(4)'s inside SO(8) and within each $SO(4)\sim SU(2)_L\times SU(2)_R$ we gauge $SU(2)_L\times U(1)_Y$. Hereafter primed objects refer to quantities and fields of the mirror sector.  Given this gauging, the $\Sigma$ field, in terms of the only physical fluctuation $\pi^{\hat{4}}=h$, reduces to
\begin{equation}\label{sigma}
\Sigma^\trans= (0,0,0,s_h,0,0,0,c_h),
\end{equation}
where $s_h \equiv \sin(h/f)$ and $c_h \equiv \cos(h/f)$. While the $Z_2$ is evident in the exchange of the two SO(4)'s inside SO(8), it acts non-linearly on the pNGBs. Indeed it can act on $\Sigma$ as
\begin{equation}
Z_2: ~~\Sigma \to R \cdot \Sigma, \quad\quad R = \left(\begin{array}{c|c} 0_4 & 1_4 \\ \hline 1_4 & 0_4\end{array} \right).
\end{equation}
The physical mode $h$ shifts by discrete values under this symmetry
\begin{equation}\label{shift}
Z_2:~~ h \to  -h + f\frac{\pi}{2}\,\quad \leftrightarrow\, \quad  s_h\to c_h.
\end{equation}
%

\subsection{Representations of fields}

Given the global symmetries in the model, we can expect the presence of resonances charged under all of them.  In this work, however, we will only be interested in those that couple directly to SM fields.  The couplings comes from the partial compositeness mechanism which couples ``elementary'' quarks to ``composite'' resonances via a linear mixing term.\footnote{Hereafter, the word ``elementary'' will refer to $q_L$ and $u_R$ fields that appear in $\mathcal{L}_{\rm mix}$.  In this basis they are neither mass eigenstates nor eigenstates of the SM gauge groups, but rather the ``elementary'' SM gauge group.  In the mass basis both gauge fields and fermions are partially composite, in general.}  This fixes the quantum numbers of the composite operators
\begin{equation}
 \mathcal{L}_{\rm mix} \sim y_L f\; \bar{q}_L \cdot \Psi + y_R f\; \bar{u}_R \cdot \Psi + (\mathrm{mirror}). \nn
\end{equation}
The elementary fields are in the usual representations of the (elementary) SM and likewise for the mirror fields.  Table~\ref{tab:reps} summarizes the possible irreducible representations of SO(8) for SM fields, mirror fields, and the composite resonances, as well as their decompositions under the relevant subgroups.  From Table~\ref{tab:reps}, it is clear that $q_L$ (and $q_L'$) can only be embedded in the \textbf{8}, while the right-handed quarks and their mirror partners have several options, namely the \textbf{1}, the \textbf{28}, or the \textbf{35}.  The gauge fields are in the adjoint representation and couple to composite vectors in the \textbf{28}.

The SM gauge fields acquire the typical masses proportional to the scale of electroweak symmetry breaking, $v$.  The mirror gauge fields, on the other hand, are not inside of SO(7) and acquire masses proportional to the goldstone scale, $f$, instead.  We expect a spectrum of the form
\begin{equation}
\begin{aligned}
 m_W \sim g v,   &\quad& m_{W'} \sim g f,   &\quad& m_\rho \sim \mathcal{O}(1-4\pi) f, \\
 m_t \sim y_t v, &\quad& m_{t'} \sim y_t f, &\quad& m_\Psi \sim \mathcal{O}(1-4\pi) f.
\end{aligned}
\end{equation}

\begin{table}[h!]
\begin{center}
\begin{tabular}{ c|c c c c }
 SM    & SO(8)  &  SO(7)  &  SO(4) $\times$ SO(4)$'$ & SU(3)$_c$ $\times$ SU(3)$_c' \times Z_2$\\
 \hline \hline
 $q_L$ & - & - & (\textbf{4},\textbf{1}) & (\textbf{3},\textbf{1})\\
 $u_R$ & - & - & (\textbf{1},\textbf{1}) & (\textbf{3},\textbf{1})\\
 $W$ & - & - & (\textbf{6},\textbf{1}) & (\textbf{1},\textbf{1})\\
 \hline \hline
 Mirror    & SO(8)  &  SO(7)  &  SO(4) $\times$ SO(4)$'$ & SU(3)$_c$ $\times$ SU(3)$_c' \times Z_2$\\
 \hline \hline
 $q'_L$ & - & - & (\textbf{1},\textbf{4}) & (\textbf{1},\textbf{3})\\
 $u'_R$ & - & - & (\textbf{1},\textbf{1}) & (\textbf{1},\textbf{3})\\
 $W'$ & - & - & (\textbf{1},\textbf{6}) & (\textbf{1},\textbf{1})\\
 \hline \hline
  Resonances    & SO(8)  &  SO(7)  &  SO(4) $\times$ SO(4)$'$ & SU(3)$_c$ $\times$ SU(3)$_c' \times Z_2$\\
 \hline \hline
 $\Psi_L$ &  \textbf{8} &  \textbf{7} $\oplus $ {\bf 1} & ({\bf 4,1}) $\oplus$ ({\bf 1,4}) &  (\textbf{3},\textbf{1}) $\oplus$ (\textbf{1},\textbf{3})\\
 $\Psi_R$ &  \textbf{1} &  \textbf{1}  & (\textbf{1},\textbf{1}) & (\textbf{3},\textbf{1}) $\oplus$ (\textbf{1},\textbf{3}) \\
 $\Psi_R$ &  \textbf{35} &  $\mathbf{27}\oplus \mathbf{7} \oplus \mathbf{1}$ & $(\mathbf{9,1})\oplus (\mathbf{1,9}) \oplus (\mathbf{4,4}) \oplus (\mathbf{1,1})$ & (\textbf{3},\textbf{1}) $\oplus$ (\textbf{1},\textbf{3})\\
 $\Psi_R$ & \textbf{28}  &  $\mathbf{21}\oplus \mathbf{7}$ &  $ (\mathbf{6,1}) \oplus (\mathbf{1,6}) \oplus (\mathbf{4,4}) $ &  (\textbf{3},\textbf{1}) $\oplus$ (\textbf{1},\textbf{3})\\
 $\rho$ & \textbf{28}  &  $\mathbf{21}\oplus \mathbf{7}$ &  $ (\mathbf{6,1}) \oplus (\mathbf{1,6}) \oplus (\mathbf{4,4}) $ &  (\textbf{1},\textbf{1})\\
 \end{tabular}
\caption{\label{tab:reps} 
Possible representations of the resonances.  Note that the composite fermions are charged under SU(3)$\times$SU(3)$'$ in a $Z_2$-invariant way.  Also note that the SM and mirror fields are embedded in incomplete representations.  The visible sector resonances are singlets of SO(4)$'$, {\it e.g.} the (\textbf{9},\textbf{1}), and likewise the mirror resonances are singlets of SO(4).}
\end{center}
\end{table}

\subsection{The non-linear $\sigma$-model} 

The \textbf{7} of SO(7) of pNGBs contains a \textbf{4} under the visible SO(4), while the other 3 pNGB's form a broken multiplet of the SM$'$ and are eaten by the mirror $W'^\pm, Z'$.\footnote{Note that under $Z_2$ the mirror photon remains massless.  One possibility to remove the mirror photon is to break the $Z_2$ in hypercharge by not gauging the mirror hyperchange~\cite{Craig:2015pha,Barbieri:2015lqa}.} Using the basis of appendix~\ref{sec:app}, the low energy non-linear $\sigma$-model is given by
\begin{equation}
\mathcal{L}= \frac{f^2}{2}(D_\mu \Sigma)^\trans D^\mu \Sigma,
\end{equation}
where $D_\mu = \partial_\mu - i g (A_\mu^a T^a_L + A_\mu^{'a} T_{L}^{'a}) $ with $T_{L}$ and $T'_{L}$ as the generators of SU(2)$_L$ and SU(2)$_L'$.  As previously noted, there is a relation between the masses of gauge bosons and their mirror partners,
\begin{equation}\label{mw-masses}
  m_W^2(h) = \frac{g^2 f^2}{4} s_h^2,
  \quad\quad\quad
  m_{W'}^2(h) = \frac{g^2 f^2}{4} c_h^2.
\end{equation} 
This expression for $m_{W}$ fixes the value of $\langle h\rangle$,
\begin{equation}\label{VEV}
v = f \sin \left(\frac{\langle h \rangle}{f}\right).
\end{equation}
We move onto the fermion sector which, given the size of the top Yukawa, gives the leading contribution to the Higgs potential.  The lagrangian of the top sector is
\begin{equation}\label{nlsm}
  \mathcal{L}= \bar{q}_L i \slashed{D} q_L + \bar{u}_R i \slashed{D} u_R + y_t f (\bar{q}_L^{\mathbf{8}})^i \Sigma_i u_R^{\mathbf{1}} + \mathrm{h.c.} + (\mathrm{mirror}).
\end{equation}
In order to write down the Yukawa-like term in the above lagrangian we have assigned the \ga  quarks to representations of SO(8). 
The notation in eq.~\eqref{nlsm} means that $q_L \in \mathbf{8}$ and $u_R\in \mathbf{1}$.  The field $u_R$ is shown in the \textbf{1} but as shown in Table~\ref{tab:reps} other representations can be used.  The embeddings are
\begin{equation}
(q_L^{\mathbf{8}})^i= \frac{1}{\sqrt{2}}\left(i b_L, b_L, i t_L, -t_L, 0, 0, 0, 0 \right)^i, \quad u_R^\mathbf{1}=u_R.
\end{equation} 
From eq.~\eqref{nlsm}, the top and its mirror partner have masses
\begin{equation}\label{mt-masses}
  m_t(h)=\frac{y_t f s_h}{\sqrt{2}},
  \quad\quad\quad
  m_{t'}(h)=\frac{y_t f c_h}{\sqrt{2}}.
\end{equation}
The ratio of these, $m_{t'} / m_t = c_h / s_h$, is typical of the mirror sector.

From eqs.~\eqref{mw-masses} and~\eqref{mt-masses}, it is possible extract the Higgs couplings to SM vectors $V$, SM fermions $f$, mirror vectors $V'$, and mirror fermions $f'$.  Normalizing the couplings to the values they have in the ordinary (unmirrored) SM, they read
\begin{equation}\label{couplings}
\begin{aligned}
  c_{hVV}   &= \sqrt{1-v^2/f^2},            &\quad\quad c_{hff}  &= \sqrt{1-v^2/f^2},\\
  c_{hV'V'} &= -\sqrt{1-v^2/f^2}(g'^2/g^2), &\quad\quad c_{hf'f'}&= -(v/f) (y'/y),
\end{aligned}
\end{equation}
where $g'$ and $y'$ denote the mirror gauge couplings and mirror Yukawa couplings, respectively, and show the effect of $Z_2$-breaking in the couplings.  The Higgs couplings to SM particles are of the usual form as in standard CH models and they are induced by the non-linearities of the $\sigma$-model.  Notice that there is a universal rescaling for the couplings of both the SM vectors and the SM fermions to the Higgs~\cite{Burdman:2014zta}.

\section{Ingredients for minimal tuning}\label{sec:tuning}

When radiatively generating the Higgs potential  there are two sources of tuning, obtaining the correct vacuum $v$ and obtaining the correct Higgs mass.  In realistic composite models, getting the correct vacuum requires a separation of scales, $v \ll f$; this tuning is always present.  Tuning in the Higgs mass is model dependent and is often worse and/or requires light top partners~\cite{Panico:2012uw}.  A natural model should then aim to tune the Higgs mass no more than the vacuum expectation value (VEV).  This scenario is called \textit{minimal tuning}
\begin{equation}\label{tuningminimal}
\Delta|_{\rm minimal} = \frac{f^2}{v^2}.
\end{equation}
The Higgs potential of CTH can satisfy minimal tuning, without the need for light colored top partners, provided some important ingredients are included.  Section~\ref{sec:ewsb} presents a systematic discussion, but here we highlight the two most important aspects.  For illustration we use the simple, though incomplete, non-linear $\sigma$-model of the previous section.

From eq.~\eqref{nlsm}, the Coleman-Weinberg potential is~\cite{Coleman:1973jx}
\begin{equation}\label{potnl}
V(h)_{\mathrm{nl}\sigma m}=\frac{N_c y_t^4 f^4}{64\pi^2} 
\left[c_h^4\log\left(\frac{2 \Lambda^2}{y_t^2 f^2 c_h^2}\right)+s_h^4\log\left(\frac{2 \Lambda^2}{y_t^2 f^2 s_h^2}\right)\right] .
\end{equation}
Due to the $Z_2$ invariance, the minimum is at $\langle h \rangle/f = \pi/4$, which is unviable for phenomenology.  On the other hand, the $Z_2$ symmetry ensures that the terms quadratically divergent in the cut-off $\Lambda$ are absent because they are proportional to $y_t^2 \Lambda^2 (s_h^2 + c_h^2)$ which is accidentally SO(8) invariant.

Additionally, the overall scale of the potential is suppressed because it is generated at $\order(y_t^4)$ the order at which the $Z_2$ no longer results in accidental SO(8) invariance. In order to attain $\langle s_h \rangle \ll 1$ the $Z_2$ symmetry needs to be broken. Assuming that the $Z_2$ breaking terms have the same parametric dependence on $y_t$ and $f$, the potential becomes
\begin{equation}\label{perfect}
V(h) = V(h)_{\mathrm{nl}\sigma m} + \frac{N_c y_t^4 f^4}{32\pi^2} b\,s_h^2 .
\end{equation}
If $b$ is a model-dependent $\order(1)$ coefficient, electroweak symmetry is broken with minimal tuning.  The Higgs mass is
\begin{equation}\label{mhperfect}
m_h^2 \simeq  \frac{N_c}{2\pi^2} \frac{m_t^2 m_{t'}^2}{f^2} \left[ \log \left(\frac{\Lambda^2}{m_t^2}\right) +\log \left(\frac{\Lambda^2}{m_{t'}^2}\right) + \order(1)\right],
\end{equation}
where $m_t$ and $m_{t'}$ are the top and mirror top masses, respectively, and $\Lambda$ is the scale where resonances will appear. In this case, minimal tuning is achieved, independent of $\Lambda$.  This example shows that the basic ingredients for a minimally tuned CTH model, without light colored top partners, are
\begin{itemize}
\item an overall scale of the potential proportional to $y_t^4 f^4$.
\item $Z_2$-breaking terms of the same numerical size of the $Z_2$-preserving ones.
\end{itemize}
In the next section we systematically study CTH for several representations of the composite fermions and different patterns for the breaking of $Z_2$.

\section{The breaking of $Z_2$ and electroweak symmetry}\label{sec:ewsb}

As we will soon see, generating the correct Higgs potential relies on breaking $Z_2$ at the right order in partial compositeness couplings $y_L$ and $y_R$.  We start by considering these couplings, especially those of the top sector, which usually give the largest contribution to the Higgs potential,
\begin{equation}\label{pc}
  \mathcal{L}= y_L f \bar{q}_L U \Psi + y_R f \bar{u}_R U \Psi + \mathrm{h.c.} + \mathcal{L}_{\rm comp}(\Psi, U, m_\Psi) + \text{mirror}.
\end{equation}
There are several important consequences that already follow from partial compositeness.  First, the parameters $y_L$ and $y_R$ break the global symmetries.  This implies that in the limit $y_{L,R} \to 0$ the Higgs potential vanishes and that the contributions start at order $y^2$ (hereafter we power count in $y \sim y_L \sim y_R$).  With the addition of the mirror sector, contributions that are $Z_2$ symmetric will start at order $y^4$.

There are several possible functional forms for the Higgs potential, which are determined by the elementary quark embeddings; a discussion of the different expressions is presented in appendix \ref{sec:app}. In this section we closely follow\cite{Giudice:2007fh,Panico:2012uw}.  In the cases of interest, the 1-loop Higgs potential generated by the top sector is
\begin{equation}\label{THpotentials}
V(h)_{\rm TH} \simeq  \frac{N_c}{16\pi^2} (y f)^{2n} m_\Psi^{2(2-n)} \bigg[ a F_{Z_2} (h/f) + b\,  F_{\slashed{Z}_2} (h/f)  \bigg], 
\quad\quad n = 1,2
\end{equation}
where for a given function, $F$, subleading terms in $y$ have been dropped.  The functions $F_{Z_2}$ and $F_{\slashed{Z}_2}$ specify the $Z_2$-preserving and $Z_2$-breaking parts of the potential, respectively.  For illustration one can consider $F_{Z_2} \sim s_h^2 c_h^2$ and $F_{\slashed{Z}_2} \sim s_h^2$.

As eq.~\eqref{pc} suggests, the Yukawa couplings for the quarks are
\begin{equation}\label{THpotentialsy}
  y_{\mathrm{SM}} \simeq y^{k} \frac{f^{k-1}}{m_\Psi^{k-1}}
\quad\quad k=1,2.
\end{equation}
Different quark representations provide different values of $n$ and $k$ which are summarized in Table~\ref{tab:functions}.  The case of $k=1$ versus $k=2$ simply reflects whether $u_R$ is fully composite or not, which can be realized for $u_R$ in the total singlet, given that this embedding does not break the global symmetries and hence the natural size of the elementary-composite mixing is $y_R f\sim m_\Psi$.   That $n=1$ for the \textbf{35} is a peculiarity of the \textbf{35} (see Table~\ref{tab:reps} and the discussion in~\cite{Mrazek:2011iu}).

\begin{table}[h]
  \begin{center}
    \begin{tabular}{|ll|cc|cc|} \hline
      & &  $n$ & $k$ & $V(h)_{\rm TH}$ & $y_{\rm SM}$  \\ \hline \hline
      $q_L^\mathbf{8}$ & $u_R^\mathbf{1}$ &  2 & 1 & $\sim y^4 f^4$          & $y$             \\
      $q_L^\mathbf{8}$ & $u_R^\mathbf{28}$ & 2 & 2 & $\sim y^4 f^4$          & $y^2(f/m_\Psi)$ \\
      $q_L^\mathbf{8}$ & $u_R^\mathbf{35}$ & 1 & 2 & $\sim y^2 f^2 m_\Psi^2$ & $y^2(f/m_\Psi)$ \\
      \hline
    \end{tabular}
    \caption{\label{tab:functions} \small Values of $n$ and $k$ for several representations of right-handed quarks as described in eqs.~\eqref{THpotentials} and~\eqref{THpotentialsy}.}
  \end{center}
\end{table}

The most favorable case is when $n=2$ and $k=1$ \textit{and} the values of $a$ and $b$ both $\sim \order(1)$.  With this choice and with $a$ and $b$ of the same size, the Higgs mass is not sensitive to $m_\Psi$ upon substituting $y_t$ into eq.~\ref{THpotentials}.  We elect to focus on this case in which $q_L \in \mathbf{8}$ and $u_R \in \mathbf{1}$ for which section~\ref{sec:concrete} presents concrete models.

Then the Higgs potential is
\begin{equation}\label{potential-ref}
  V(h)_{\rm TH}\simeq \frac{N_c}{16\pi^2} y_t^4 f^4 \big(-a s_h^2 c_h^2 + \lambda b s_h^2  \big),
\end{equation}
where $a \sim \order(1)$ and describes the $Z_2$-symmetric contribution, while $\lambda b$ describes the size of the $Z_2$-breaking contribution where $b \sim \order(1)$ and $\lambda$ is introduced to parameterize scaling deviations from $\order(1)$.  The electroweak VEV and the Higgs mass demand
\begin{equation}\label{minimum}
  \frac{v^2}{f^2}= s_h^2 = \frac{a-\lambda b}{2a}, 
  \quad\quad\quad
  m_h^2 \sim a\, \frac{N_c}{2\pi^2} y_t^4 v^2.
\end{equation}
The amount of cancellation needed to realize the Higgs VEV can be larger than prescribed by minimal tuning.  Rearranging eq.~\eqref{minimum}, one sees that it is required that
\begin{equation}
  a-\lambda b \sim 2 a \frac{v^2}{f^2}.
\end{equation}
In the case of large $\lambda \gg \order(1)$, since $a$ is determined by the $Z_2$-symmetric potential, one needs to first tune $b$ to compensate for $\lambda$ and then tune $a$ and $\lambda b$ to get the right VEV.  This results in a double tuning and generically predicts
\begin{equation}\label{tuning-generic}
  \Delta \sim \frac{f^2}{v^2}\, \lambda.
\end{equation} 
Notice that in the opposite case $\lambda \ll \order(1)$, it is $a$ that must be tuned.  Enforcing cancellations to tune $a$, however, will spoil the agreement with the Higgs mass \eqref{minimum}, as it will turn out to be too light.

The reader may observe that the logarithmic dependence on $h$ in the terms in eq.~\eqref{potnl} has been neglected in favor of the simple functional form $a s_h^2 c_h^2$.  The difference from the $h$ dependence is subleading (see appendix~\ref{sec:app} which obtains the leading behavior from a spurion analysis) and is small enough that we can still obtain parametric estimates for the Higgs mass and tuning.  In the concrete models of section~\ref{sec:concrete} we keep track of the logarithmic effects (which basically arise from the running from the threshold to the weak scale).

\subsection{$Z_2$-breaking in the top sector}

Breaking $Z_2$ only within the top sector tends to spoil the TH mechanism almost completely, effectively reducing the CTH framework to the standard class of CH models.  In the top sector there are two possibilities to break $Z_2$: $(i)$ an $\order(1)$ breaking in the composite sector via $m_\Psi \neq m_{\Psi'}$ (especially among the SO(7)-invariant parameters) or $(ii)$ breaking in the elementary sector via $y \neq y'$.  There is additionally an exception which occurs when $u_R \in \mathbf{35}$.

For both these cases the Higgs potential takes the form
\begin{equation}\label{Z2breakingtop}
  V(h)_{\rm TH}\simeq  \frac{N_c}{16\pi^2} \bigg[-a y^4 f^4 s_h^2 c_h^2+ b\, y^2 f^2 m_\Psi^2 s_h^2  \bigg].
\end{equation}
We see that a suppression of $b$ of the size $\sim y^2 f^2 / m_\Psi^2$ is required.  Assuming $b$ is of $\order(1)$, this leads to a Higgs mass and tuning of
\begin{equation}
  m_h^2 \simeq \frac{a N_c y_t^4 v^2}{2\pi^2},
\quad\quad\quad
\Delta  \simeq \frac{f^2}{v^2}\frac{m_\Psi^2}{y^2 f^2}.
\end{equation}
While the Higgs mass does not depend on the scale of colored fermion partners, $m_\Psi$, the VEV in fact does.  The departure from minimal tuning as a result of the $Z_2$-breaking grows as the ratio of the mirror top to the colored top partner.  In absolute terms, the tuning grows proportionally to the mass of the top partner
\begin{equation}\label{tun}
  \Delta \sim \frac{f^2}{v^2}\frac{m_\Psi^2}{m_{t'}^2}\sim \frac{m_\Psi^2}{m_t^2}.
\end{equation}
Therefore, the naturalness of the model implies a light colored particle $\Psi$ and truly nothing is gained relative to the usual tuning of CH models.  Even the phenomenology of CTH in this case is almost identical to standard CH models.\footnote{Readers may note a difference between eqs.~\eqref{predictionsCH} and~\eqref{tun}.  That the Higgs mass is independent of $m_\Psi$ can occur in standard CH models, at the price of tuning, when $t_R$ is a total singlet (see section 5.3 of~\cite{Marzocca:2012zn}).}

The difference between breaking in the composite sector as opposed to the elementary sector comes down to the expected size of symmetry violating parameters.  When broken in the elementary-composite mixings, the $Z_2$-breaking parameter goes like $b \sim (y^2 - y'^2)/y^2 \times (f^2/m_\Psi^2)$ which can naturally reduce to $b \sim y^2 f^2 / m_\Psi^2$ by tuning $y$ and $y'$ against each other.  Breaking in the composite sector entails a mass splitting between composite parameters, which without extra assumptions would be of the order $\delta m \equiv | m_\Psi - m_{\Psi'} | \sim m_\Psi$, making it difficult to realize the needed cancellation.

\subsubsection*{Almost $Z_2$-symmetric composite sector}

Although it may look unappealing, breaking the symmetry in the strong sector could be reconciled with minimal tuning in the case when the composite sector is almost $Z_2$-symmetric.  As the potential is
\begin{equation}\label{almostZ2}
  V(h)_{\rm TH}\simeq  \frac{N_c}{16\pi^2} \bigg[-a y^4 f^4 s_h^2 c_h^2+ b\, y^2 f^2 (m_\Psi^2 -m_{\Psi'}^2) s_h^2  \bigg] ,
\end{equation}
one can see that taking $m_\Psi^2 - m_{\Psi'}^2$ to zero restores a symmetry so that the model can be considered technically natural.  This was also the case when breaking with $y^2 - y'^2$.  Schematically the tuning can be
\begin{equation}
  \Delta \sim \frac{f^2}{v^2}\frac{m_\Psi^2-m_{\Psi'}^2}{y_t^2 f^2}
  \quad \stackrel{\mathrm{almost\ Z_2}}{\sim} \quad
  \frac{f^2}{v^2}.
\end{equation}
If the breaking is stable, the model has minimal tuning and the Higgs mass not sensitive to colored top partners.  The only condition is maintaining a small difference in masses.  Despite the fact that we do not know any mechanism for generating such a mass difference in the composite sector, we do not disregard this as a possibility.

\subsubsection*{$Z_2$-breaking in right-handed sector with $u_R\in \mathbf{35}$}

For the \textbf{35}, unlike for the other representations, the leading contributions to the Higgs potential go like $y^2 c_{2h}^2$ and $y^4 s_h^2$ (see appendix~\ref{sec:app} for a more complete list).  From this one can see that the contribution of the SM particles and the mirror particles to the Higgs potential is not accidentally SO(8) invariant at leading order, $\order(y^2)$, even with the $Z_2$ symmetry.  This means the $Z_2$-symmetric contribution is $\order(y^2)$ rather than $\order(y^4)$.

On the other hand, the $y^2$ term is already $Z_2$-invariant without the addition of the mirror particles such that one must go to $\order(y^4)$ to break $Z_2$.  Thus the relative contributions from the $Z_2$-breaking and $Z_2$-preserving terms are reversed relative to the other representations.  The potential, however, is formally equivalent to eq.~\eqref{Z2breakingtop} (renaming the terms $s_h^2c_h^2 \leftrightarrow s_h^2$) so the same predictions regarding tuning apply.

\subsection{$Z_2$-breaking in the lighter quarks}

Another possibility is to preserve $Z_2$ in the top sector, but break it among the bottom quark or the quarks of the first two generations.  Numerically the only couplings that can be relevant for generating a sufficiently large breaking of $Z_2$ are the bottom and charm Yukawa, $y_b$ and $y_c$ (hereafter we refer to the bottom or charm as the ``lighter quarks'').\footnote{For a discussion of the impact of the lepton sector and how it can help to raise the top partner mass in non-twin CH models see~\cite{Carmona:2014iwa}.}  The breaking will produce a potential of the form
\begin{equation}\label{charm-bottom}
  V(h)_{\rm TH}\simeq  \frac{N_c}{16\pi^2} \bigg[-a y_t^4 f^4 s_h^2 c_h^2 +b\, y_q^2 f^2 m_\Psi^2 s_h^2   \bigg],
\end{equation}
where $y \sim y_{L} \sim y_{R}$ is the elementary-composite coupling of the lighter quark. For simplicity, we do not consider the mirror contributions which are proportional to $-y'^2 s_h^2 $. If the $q_R$ is embedded in the \textbf{1}, then the $Z_2$-breaking term is too small; the parameter $a$ requires tuning to realize the EWSB and as a result the Higgs mass will be extremely light.  With the $q_R \in \mathbf{28}$, however, the elementary-composite parameter $y$ is proportional to
\begin{equation}
  y^2 \sim y_{\rm SM} \frac{m_{\Psi}}{f}. \nn
\end{equation}
\textit{i.e.} $k=2$ from Table~\ref{tab:functions}.  The potential in terms of the known Yukawa coupling is
\begin{equation}\label{lightq-estim}
  V(h)_{\rm TH}\simeq  \frac{N_c}{16\pi^2} \bigg[ -a y_t^4 f^4 s_h^2 c_h^2 + b\, y_q^{\rm SM} f m_\Psi^3 s_h^2 \bigg], \quad q=b,c.
\end{equation}
If $m_\Psi \sim (y_t^{4}/y_{b,c})^{1/3}f$, then the terms are comparable, with $a$ and $b$ of size $\order(1)$.  As a result the tuning is minimal provided the fermionic resonances are at a scale $\sim 4f$ and $\sim 7f$ for the bottom and charm partners, respectively
\begin{equation}\label{tuning-lightquarks}
  m_h^2 \simeq \frac{a N_c y_t^4 v^2}{2\pi^2},
  \quad\quad
  \Delta\big|_{\slashed{Z_2} -\mathrm{bottom}} \sim \frac{f^2}{v^2}\left(\frac{m_\Psi}{4f}\right)^3, 
  \quad\quad
  \Delta\big|_{\slashed{Z_2} -\mathrm{charm}} \sim \frac{f^2}{v^2}\left(\frac{m_\Psi}{7f}\right)^3.
\end{equation}
We see the Higgs mass itself that is not sensitive to $m_\Psi$.

\subsection{$Z_2$-breaking in the gauge sector}

Another possible scenario involves the SU(2) gauge coupling $g$, which is only marginally smaller than $y_t$.  In particular we work in the exact $Z_2$ limit in the quark sector, allowing for the breaking of $Z_2$ by the gauge fields. In this case the potential is
\begin{equation}\label{pot-gauge}
  V(h)_{\rm TH} \simeq \frac{1}{16\pi^2} \bigg[-a N_c y_t^4 f^4 s_h^2 c_h^2 + b \frac{9}{4}\frac{g^2 m_\rho^4}{g_\rho^2} s_h^2 \bigg],
\end{equation}
where $m_\rho$ is the mass of a vector composite resonance, likely a \textbf{21} of SO(7), with a coupling $g_\rho$ to the composite sector.  Despite the fact that this potential contribution from the gauge fields is known in composite Higgs models, we review relevant details for convenience.  The SU(2) contribution is given  by \cite{Agashe:2004rs}
\begin{equation}
  V(h)_{\rm SU(2)} = \frac{9}{2}\int \frac{d^4p}{(2\pi)^4} \log \bigg[1 + \frac{g^2}{g_\rho^2}\frac{s_h^2}{2} F(p^2) \bigg],
\end{equation}
where $F(p^2)$ is a form factor with poles at resonance masses, $\sim m_\rho$, while the mirror contribution is given by $g\to g'$ and $s_h \to c_h$, where $g'$ is the mirror SU(2) coupling.  In the simplest case $F(p^2) \sim m_\rho^4 /(p^2(p^2-m_\rho^2))$.  In an expansion in $g/g_\rho$ the leading contributions from SU(2) read
\begin{equation}
  V(h)_{\rm  SU(2)} = \frac{9}{64\pi^2} m_\rho^4 \bigg[ b \frac{g^2}{g_\rho^2} s_h^2 + b' \frac{g^4}{g_\rho^4} s_h^4 \bigg],
\end{equation}
where the term $\order(g^4/g_\rho^4)$ is a subleading contribution to the $Z_2$-symmetric potential and is neglected as it is small compared to the top sector contribution in eq.~\eqref{pot-gauge}.  The coefficient $b$ is expected to be of $\order(1)$, $b < \log (16\pi^2/g_\rho^2)$.

Here we use $m_\rho \sim g_\rho f$ (we consider $m_\rho$ and $m_\Psi$ to be different parameters).  The gauge contribution can then be added to the $Z_2$-symmetric potential to find
\begin{equation}\label{potential-gauge-z2breaking}
  V(h)_{\rm TH} \simeq \frac{f^4}{16\pi^2} \bigg[-a N_c y_t^4 s_h^2 c_h^2 + b \frac{9}{4} g_\rho^2 g^2 s_h^2\bigg].
\end{equation}
The above expression is valid in the case of a maximal $Z_2$ breaking, $g'=0$. In the case of finite $g'$, eq.\eqref{potential-gauge-z2breaking} should be modified by making the replacement $g^2 \to g^2 - g'^2$. Minimal tuning occurs for $m_\rho \sim 4 f$.  Following the same arguments as the previous sections, the tuning and the Higgs mass are predicted to be
\begin{equation}
  m_h^2 \simeq a\frac{N_c y_t^4}{2\pi^2}v^2,
  \quad\quad
  \Delta \simeq \frac{f^2}{v^2} \left(\frac{g_\rho}{3.5}\right)^2.
\end{equation}
In this scenario the Higgs mass and tuning are not sensitive to the mass of the colored top partners.

An additional, potentially relevant, consequence of breaking $Z_2$ in the gauge sector is that two-loop effects will cause the visible and mirror top Yukawas to run differently, reintroducing a term proportional to $f^2 m_\Psi^2$ in the Higgs potential.  Indeed, over a decade of running, one can estimate the visible and mirror top Yukawa to be split by (see {\it e.g.} \cite{Craig:2015pha})
\begin{equation}
   | y_t^2 - y_{t'}^2 | \simeq y_t^2 \frac{9g^2}{64\pi^2} .
\end{equation}
Above we assume the gauge couplings maximally break $Z_2$.  This induces the second term of eq.~\eqref{Z2breakingtop} which is
\begin{equation}
  \Delta V(h)_{\rm TH} \simeq \frac{N_c}{16\pi^2} 
  \bigg[ b_{\rm 2-loop} \frac{9g^2}{64\pi^2} y_t^2 f^2 m_\Psi^2 s_h^2 \bigg] ,
\end{equation}
where $b_{\rm 2-loop}$ is an $\order(1)$ coefficient.  To maintain the tuning estimates we make, the above contribution should not dominate the $Z_2$-breaking terms in the potential, the second term of eq.~\eqref{potential-gauge-z2breaking}, which requires
\begin{equation}
  \left(\frac{m_\Psi}{4\pi f}\right)^2 \frac{N_c y_t^2}{g_\rho^2} \frac{b_{\rm 2-loop}}{b} \lesssim 1.
\end{equation}
While this condition yields an upper bound on $m_\Psi$, because $g_\rho$ can be as large as $\sim 3.5$ we find $m_\Psi \lesssim 4\pi f$ to be an appropriate estimate.  Of course it may be the case that $y_t$ and $y_{t'}$ are not exactly equal at the scale $\Lambda$, but we neglect this case as it deviates from the spirit of the twin mechanism.  We will neglect the effect of this two-loop contribution in the rest of the paper.

\section{Concrete models}\label{sec:concrete}

Motivated by the above discussion we elect to focus on two models.  In the first $Z_2$-breaking is introduced in the gauge sector and in the second it is induced in the lighter quarks.  In both cases, the Higgs potential can be parameterized as
\begin{equation}
V(h) \simeq - \alpha s_h^2 c_h^2 + \beta s_h^2.
\end{equation}
The Higgs VEV and mass are given, respectively, by
\begin{equation}\label{vevHiggs}
v= \sqrt{\frac{\alpha-\beta}{2\alpha}}f
\quad\quad
m_h^2 = \frac{8\alpha}{f^4}v^2 \left(1-\frac{v^2}{f^2}\right).
\end{equation}
As explained in section~\ref{sec:ewsb} the Higgs mass is set by the $Z_2$-symmetric piece $\alpha$.  Because $\alpha$ is already of the right order to give $m_h = 125$ GeV, the Higgs mass computation is the same for both models, as we will show below.  The computation of the $Z_2$-breaking term, $\beta$, determines the tuning for each model.

In the models under consideration, the scaling of the two terms are,
\begin{itemize}

  \item Model A: $Z_2$-breaking in gauge sector
    \begin{equation}
      \alpha \sim y_t^4 f^4,\quad \beta \sim g^2 g_\rho^2 f^4. \nn
    \end{equation}

  \item Model B: $Z_2$-breaking in lighter quarks
    \begin{equation}
      \alpha \sim y_t^4 f^4,\quad \beta \sim y_q m_\Psi^3 f. \nn
    \end{equation}

\end{itemize}
We do not consider the case of $Z_2$-breaking in the top sector, because this is typically equivalent to a standard CH model.

\subsection{Computing the Higgs mass}

We start with the computation of $\alpha$, which sets the Higgs mass via eq.~\eqref{vevHiggs}.  We consider a 2-site model with two composite fermion resonances~\cite{Panico:2011pw}, $\Psi_7$ and $\Psi_1$, which are in the \textbf{7} and the \textbf{1} of SO(7), respectively.  The elementary $q_L$ is embedded in the \textbf{8} and the $t_R$ is a total singlet (a chiral composite state).  We will show explicitly that $\alpha$ is fully calculable. The model is defined by
\begin{equation}\label{2-site}
\begin{split}
\mathcal{L} &= \bar{q}_L i \slashed{D} q_L + \bar{u}_R i \slashed{D} u_R + y_L f 
              (\bar{q}_L^{\mathbf{8}})^i (U_{iJ} \Psi_7^J + U_{i8} \Psi_1) + \mathrm{h.c.} \\
            & + \bar{\Psi} i \slashed{D} \Psi - m_1 \bar \Psi_1 \Psi_1 - m_7 \bar{\Psi}_7 \Psi_7 - m_R (\bar{\Psi}_1)_L u_R^\mathbf{1}\\
& + \mathrm{(mirror)}.
\end{split}
\end{equation}
In order to compute the Higgs potential, we can integrate out the composite sector and match to the following effective lagrangian, constructed just with the low energy fields and the $\Sigma$ in the SO(7) vacuum~\cite{Agashe:2004rs}
\begin{equation}\label{eff}
\begin{split}
\mathcal{L}_{\rm eff}&= (\bar{q}_L^\mathbf{8})_i \slashed{p}(\delta^{ij}\Pi_0^q(p) + \Sigma^i\Sigma^j \Pi_1^q(p)) 
  (q_L^\mathbf{8})_j + \bar{u}_R \slashed{p} \Pi_0^u(p) u_R 
  + (M(p)(\bar{q}_L^\mathbf{8})_i \Sigma^i u_R + \mathrm{h.c.}) +  (\mathrm{SM} \to \mathrm{SM}')\\
  &=\bar{u}_L \slashed{p}\left(\Pi_0^q(p) + \Pi_1^q(p) \frac{s_h^2}{2}\right) u_L 
  + \bar{u}_R \slashed{p} \Pi_0^u(p) u_R + \left(\frac{M(p)}{\sqrt{2}}\bar{u}_L s_h u_R  +\mathrm{h.c.}\right)+  (\mathrm{SM} \to \mathrm{SM}').
\end{split}
\end{equation}
From an explicit calculation, the form factors are
\begin{equation}
\begin{split}
&\Pi_0^q(p) = 1 - \frac{y_L^2 f^2}{p^2-m_7^2}, \quad\quad
\Pi_1^q(p) = \frac{y_L^2 f^2 (m_7^2-m_1^2)}{(p^2-m_7^2)(p^2-m_1^2)}, \\
&\Pi_0^u(p) = 1-\frac{m_R^2}{p^2-m_1^2}, \quad\quad
M(p) = - \frac{y_L m_1 m_R}{p^2-m_1^2}.
\end{split}
\end{equation}
We can compute the top mass, which is
\begin{equation}\label{eq:topmass}
m_t \simeq \frac{y_L f m_R s_h}
{\sqrt{2(m_1^2+m_R^2)\left(1+ \frac{y_L^2 f^2}{m_7^2}\left(1+ \frac{m_7^2-m_1^2}{m_1^2}\frac{s_h^2}{2}\right)\right)}},
\end{equation}
and the same for $m_{t'}$, under $s_h\to c_h$.  The 1-loop potential generated by the above lagrangian is
\begin{equation}\label{eq:pintegral}
  \begin{split}
  V(h)&= -2 N_c \int \frac{d^4p}{(2\pi)^4}\log \bigg[p^2\left(\Pi_0^q(p) + \Pi_1^q(p) \frac{s_h^2}{2}\right) \Pi_0^u(p)
    - M(p)^2 \frac{s_h^2}{2} \bigg] + (s_h \to c_h)\\
& = - N_c y_L^4 f^4 s_h^2 c_h^2 \int \frac{d^4p}{(2\pi)^4}\frac{\left(m_1^2 p^2+m_7^2 (m_R^2-p^2)\right)^2}{2 p^4 (m_7^2-p^2)^4 \left(m_1^2+m_R^2-p^2\right)^2},
\end{split}
\end{equation}
where in the second line we have performed an expansion to $\order(y_L^4)$, the first non-zero order in $y_L$.  The integration is performed using a lower limit of $m_{t'} \sim y_L f$ in order to account for the IR effects associated with the massive top and its mirror partners (see eq.~\eqref{mhperfect}).

In the limit where $y_L f \ll m_\Psi$ the mass spectrum of the resonances (and mirror resonances) is approximately SO(7)-invariant with mass eigenstates for the \textbf{7} and \textbf{1} resonances
\begin{equation}
\bar{m}_{\mathbf{7}}\simeq m_7,
\quad\quad
\bar{m}_{\mathbf{1}}\simeq \sqrt{m_1^2 + m_R^2}.
\end{equation}
When one takes the interesting limit of heavy fermionic composite parameters $\bar{m}_{\mathbf{7}}, \bar{m}_{\mathbf{1}} \gg y_L f$, (and for simplicity $m_1/m_R\simeq 1 +O(y^2_L)$), eq.~\eqref{eq:pintegral} simplifies and the expression for $\alpha$, upon imposing the top mass in eq.~\ref{eq:topmass}, is
\begin{equation}\label{eq:alphafull}
\alpha \simeq \frac{N_c y_t^4 f^4}{32\pi^2}
\bigg[ \log \left(\frac{\bar{m}_{\mathbf{1}}^2}{m_{t'}^2}\right) 
 - 5\bigg(1- \frac{4}{5}\frac{\bar{m}_{\textbf{7}}^2}{\bar{m}_{\textbf{7}}^2-\bar{m}_{\textbf{1}}^2}
 \log \left(\frac{\bar{m}_{\textbf{7}}^2}{\bar{m}_{\textbf{1}}^2}\right)\bigg) + \order\left(\frac{y_L f}{m_\Psi}\right) \bigg].
\end{equation}
Note that another possible limit is where all the fermionic composite parameters originate from a common scale $m_\Psi$ , $m_7/m_1\simeq m_1/m_R\simeq 1 +O(y^2_L)$.  In this case we have $\bar{m}_{\mathbf{7}}\simeq \bar{m}_{\mathbf{1}}/\sqrt{2}$ and eq.~\eqref{eq:alphafull} simplifies even more.  The final result for the Higgs mass is
\begin{equation}\label{HiggsMassModel}
m_h^2 \simeq \frac{N_c y_t^4 v^2}{4\pi^2}\left(1-\frac{v^2}{f^2}\right)
\left[ \log \left(\frac{\bar{m}_{\mathbf{1}}^2}{m_{t'}^2}\right) - 5\bigg(1- \frac{4}{5}\frac{\bar{m}_{\textbf{7}}^2}{\bar{m}_{\textbf{7}}^2-\bar{m}_{\textbf{1}}^2}\log \left(\frac{\bar{m}_{\textbf{7}}^2}{\bar{m}_{\textbf{1}}^2}\right)\bigg) \right].
\end{equation}
Hence, no light colored partners are required to achieve 125 GeV.  Using $m_{t'} \simeq m_t f/v$ (where $m_t(\mathrm{TeV}) \simeq 150$ GeV), we find to get $m_h = 125$ GeV we have an overall fermionic scale which is almost unconstrained, $m_\Psi \gtrsim 4 f$.

\subsection{Model A: $Z_2$-breaking in the gauge sector}

\begin{figure}
\begin{center}
\includegraphics[width=.49\textwidth]{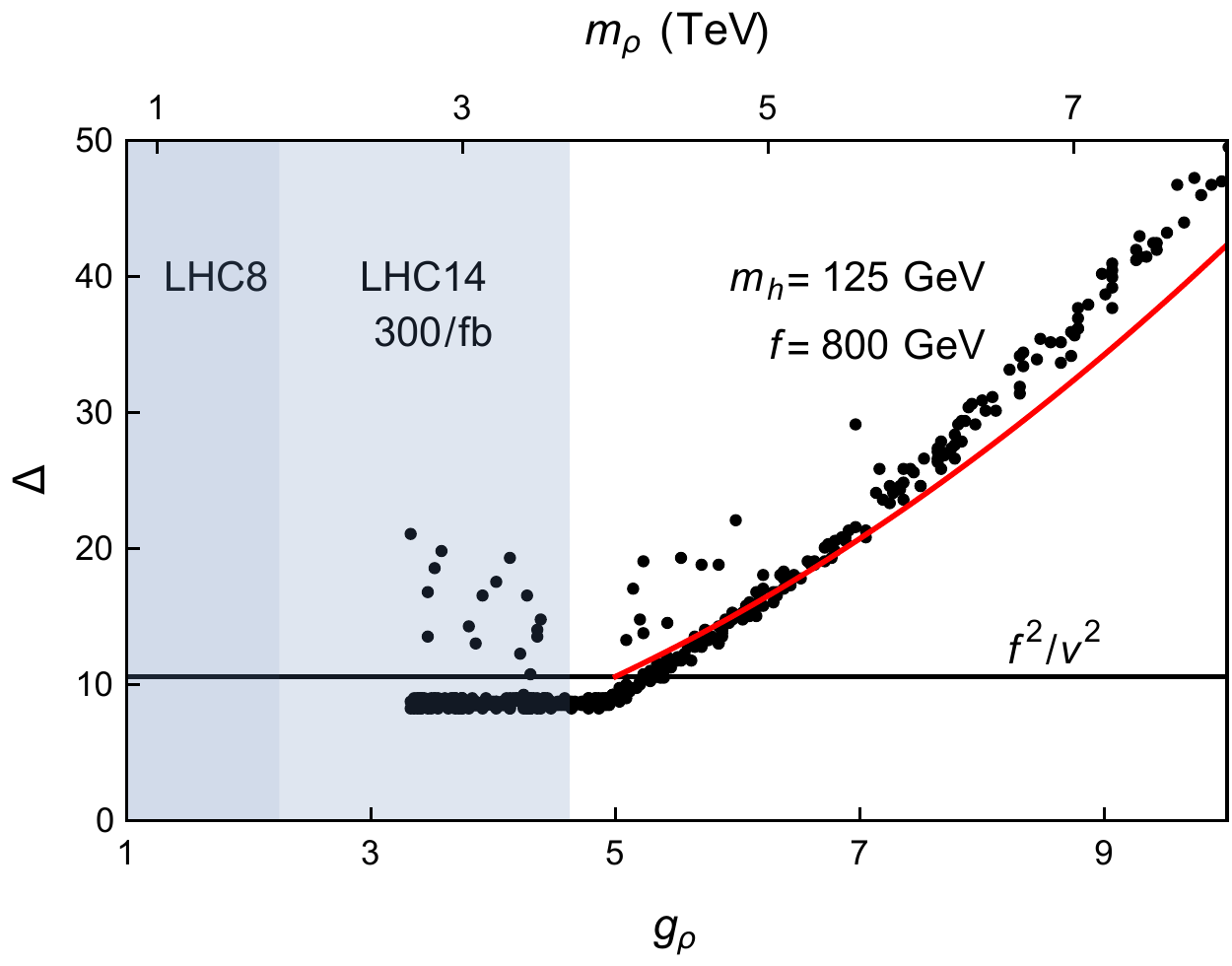}~\includegraphics[width=.49\textwidth]{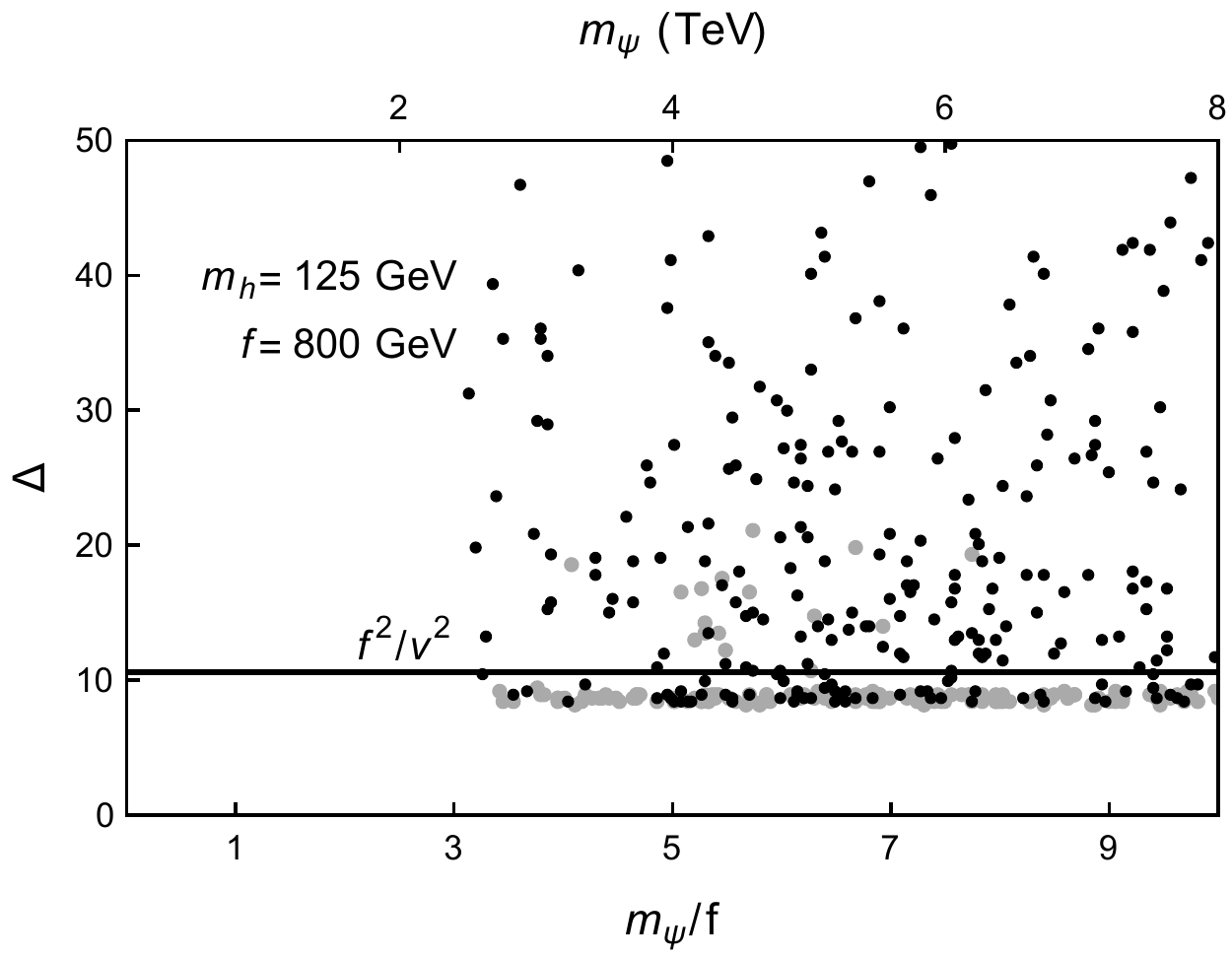}
\caption{\label{fig:tuning} Model A: $Z_2$-breaking in the gauge sector, with $f=800$ GeV and $\kappa=1$. Left: tuning versus $g_\rho$. The growth in the tuning with respect to the minimal value $f^2/v^2$ around $g_\rho\sim 5$ is due to the fact that one has to invoke a cancellation between the $g^2-g'^2$, which is taken into account in the numerical analysis.  The red line is the scaling with $g_\rho^2$ as predicted by eq.~\eqref{tunGAUGE}. Right: Tuning versus $m_\Psi/f$. We scan over a range $[0.5,10]$ TeV for composite masses and $[2,10]$ for $g_\rho$, where the top mass is fixed at $m_t(\sim \mathrm{TeV})=150$ GeV.  We defined $m_\Psi$ as the geometric average of the composite masses. The gray points correspond to values of $g_\rho$ that LHC14 will probe (left panel).}
\end{center}
\end{figure}

Next, we compute $\beta$ when $Z_2$ is broken in the SU(2) gauge fields, allowing for different $g\neq g'$ between SU(2) and its mirror. The gauge sector of the model in eq.~\eqref{2-site} is
\begin{equation}
\mathcal{L}= -\frac{1}{4}(F_{\mu\nu}^2 +{\rm mirror}) -\frac{1}{4}\rho_{\mu\nu}^2 + \frac{f^2}{4}\mathrm{Tr}[(D_\mu U)^\trans D_\mu U],
\end{equation}
where the covariant derivative is $D_\mu U = \partial_\mu U - i g A^a_\mu T_L^a U - i g' A'^a_\mu T_L^a U + i g_\rho U \rho^A_\mu T^A$ and $g'$ is the mirror SU(2) coupling. The discussion of the hypercharge formally follows the same steps.  In the 2-site model, where we have only a vector $\rho_\mu$ in the \textbf{21} of SO(7), $\beta$ is logarithmically divergent, $\beta \sim \log (\Lambda^2/m_\rho^2)$. 

From the above lagrangian we have that the mass of $\rho$, to zeroth order in the mixings, is $m_\rho^2 = g_\rho^2 f^2/2$, which leads, in the case $g'=0$, to $\beta = 9/(256 \pi^2) g^2\, g_\rho^2\, f^4 \log (\Lambda^2/m_\rho^2) + \order(g^4)$.  To render this contribution finite it is crucial to have the first coset resonance (a \textbf{7} of SO(7), dubbed $a_1$) in the low energy theory.  We then find
\begin{equation}\label{beta-gauge}
\beta = \frac{9}{64\pi^2} m_\rho^4 \frac{g^2-g'^2}{g_\rho^2}\log \frac{m_{a_1}^2}{m_\rho^2} + \order(g^4),
\end{equation}
in which the contribution from the mirror SU(2) is included.  Depending on the model we can have different predictions for $m_\rho$ and $m_{a_1}$.  If we take the model of~\cite{DeCurtis:2011yx} they are related by an extra parameter, $\kappa$ and the masses are
\begin{equation}
m_\rho^2 = g_\rho^2 f^2 \frac{1+\kappa^2}{2\kappa},
\quad\quad
m_{a_1}^2 = g_\rho^2 f^2 \frac{(1+\kappa^2)^2}{2\kappa},
\end{equation}
where $1+\kappa^2 < 16\pi^2/g_\rho^2$. 

Interestingly, the gauge contribution needed to obtain the right VEV is found for $g_\rho \sim 4$.  Depending on the value of $\kappa$ and the size of the mirror SU(2) coupling $g'$, which cannot be taken to zero to avoid massless bosons; $g_\rho$ can be even larger, $g_\rho \lesssim 6$.  It seems difficult, however, to push it as high as $4\pi$ without going into a region where $g' \approx g$.  In any case, the tuning does not scale with $m_\Psi$ as long as $g_\rho \lesssim 6$,
\begin{equation}\label{tunGAUGE}
\Delta \simeq \frac{f^2}{v^2} \bigg(\frac{g_\rho}{5}\bigg)^2.
\end{equation}
With this concrete result in hand, we can now make statements about the tuning of the model, the vector resonances, and the fermion resonances.

As shown, minimal tuning is achieved for $g_\rho \sim 4 \div 5$ which also gives the correct Higgs mass.  With a small price in tuning, $g_\rho$ can be made larger, but it cannot be made much smaller because the Higgs mass would be too small.  A full numerical computation of the tuning $\Delta$, computed with the Barbieri-Giudice measure~\cite{Barbieri:1987fn} with respect to the model parameters,\footnote{The tuning is defined using the Barbieri-Giudice measure~\cite{Barbieri:1987fn},
\begin{equation}
  \Delta = \max_{i} \left| \frac{\partial \log v^2}{\partial \log x_i} \right|,
\end{equation}
where $x_i$ are the input parameters to the theory.  We take the $x_i$'s to be $ x_i = \{ m_1, m_7, m_R, g_\rho, g' \}$. The parameter space scans are performed sampling each mass parameter uniformly between $0.5$ TeV and $10$ TeV. For each parameter space point, the tuning, the VEV, and the Higgs mass, are computed.  The parameter $y_L$ is selected to get $m_t(\mathrm{TeV}) = 150$ GeV, while scanning over $g'$ we required the mass of the mirror vectors to be sufficiently large ($m_{W'}>125$ GeV).  Points drawn are required to fall within $ 240~\mathrm{GeV} < v < 250~\mathrm{GeV}$, $  123~\mathrm{GeV} < m_h < 127~\mathrm{GeV}$, otherwise they are rejected.} 
is shown in figure~\ref{fig:tuning}.  These plots use $f \simeq 800$ GeV, a value that will be probed by precision measurements at LHC14 with $300$ fb$^{-1}$~\cite{ATLAS-Collaboration:2012jwa,CMS-projections}.  The left panel shows that our estimate in eq.~\eqref{tunGAUGE} is respected.  The right panel shows tuning versus $m_\Psi/f$ (here $m_\Psi$ is defined as a geometric average of the composite parameters).  We see that one can have both low tuning, $\sim 5 \div 10 \%$, and heavy colored top partners.

Using a simple rescaling of~\cite{Contino:2013gna} (shown in figure~\ref{fig:tuning}) we see that the LHC will probe some of the low tuning region via direct searches for the $\rho$, but that model points \textit{with minimal tuning} will \textit{survive} constraints from LHC14.

If the breaking in the gauge sector is entirely due to hypercharge, while SU(2)$'$ is exactly $Z_2$-invariant, the estimate on the natural size of $g_\rho$ can be lifted by a factor $\sqrt{3} g/g_Y \sim 2 \div 3$.\footnote{We became aware of this possibility after a discussion with the authors of~\cite{Barbieri:2015lqa}. In the limit where the mirror hypercharge is not gauged, $g_Y \neq g_Y'\equiv 0$, the massless mirror photon is absent from the spectrum.}

\subsection{Model B: $Z_2$-breaking in the lighter quarks}

In our second model the $Z_2$-breaking terms come entirely from the lighter quarks, while the top and gauge sectors are fully $Z_2$-symmetric.  The contribution to the potential, as discussed in eq.~\ref{charm-bottom}, is proportional to $y^2 f^2 m_\Psi^2$.  Recall that $y$ is the elementary-composite mixing and its relation to $y_q=\sqrt{2}m_q/v,\, q=b,c,$ depends on the representations of $q_L$ and $q_R$.  Referring to Table~\ref{tab:functions}, only $k=2$ can satisfy minimal tuning for the light quarks.

At a practical level, this leads us to embed the right-handed bottom (and charm) in the \textbf{28},
\begin{equation}
\begin{split}
\mathcal{L} &= \bar{q}_L i \slashed{D} q_L + \bar{q}_R i \slashed{D} q_R 
+ y_R f (\bar{q}_R^{\mathbf{28}})^{ij} ( U_{jJ} U_{iL} \Psi_{21}^{JL} + U_{i8}U_{jJ} \Psi_7^J) 
+y_L f (\bar{q}_L^{\mathbf{8}})^i (U_{iJ} \Psi_7^J + U_{i8} \Psi_1) + \mathrm{h.c.} \\
& + \bar{\Psi} i \slashed{D} \Psi  -\tilde{m}_1 \bar \Psi_1 \Psi_1 - \tilde{m}_{21} \bar \Psi_{21} \Psi_{21} - \tilde{m}_7 \bar{\Psi}_7 \Psi_7 ,
\end{split}
\end{equation}
where we neglected the mirror contribution. This lagrangian for the bottom (or charm) sector is similar to that of the top sector in eq.~\eqref{2-site}, but also has a resonance in the \textbf{21} of SO(7).  To remove any potential confusion, composite mass parameters of this sector are labeled with tildes.  The computation for $\alpha$ follows the same steps for computing $\beta$.  Computing the Coleman-Weinberg potential we find, to leading order in $m_q/v$
\begin{equation}\label{alpha-q}
\beta \simeq  \frac{N_c}{8\pi^2} \frac{m_q f}{x_q v}\tilde{m}_7 \bigg( \tilde{m}_1^2 \log\frac{\tilde{M}^2}{\tilde{m}_1^2} + x_q^2 \tilde{m}_{21}^2 \log \frac{\tilde{M}^2}{\tilde{m}_{21}^2} - (1+x_q^2) \tilde{m}_7^2 \log \frac{\tilde{M}^2}{\tilde{m}_7^2} \bigg),
\end{equation}
where $x_q = y_R/y_L$ is the ratio of elementary-composite couplings
\begin{equation}
y_L \simeq \sqrt{\frac{2 \tilde{m}_7 m_q}{x_q f v}},
\quad\quad
y_R \simeq \sqrt{\frac{2 x_q \tilde{m}_7 m_q}{f v}},
\end{equation}
and $\tilde{M}$ is the cut-off scale.  The introduction of this scale is necessary because like the 2-site contribution to $\beta$ in Model A, $\beta$ is logarithmically divergent in the effective description.  For the estimates we present we replace the cut-off with $\tilde{M}$ which represents the next layer of resonances that would be present in a 3-site or 5D model.  Eq.~\eqref{alpha-q} shows that the scaling estimated in eq.~\eqref{lightq-estim} is respected.

Assuming a common mass scale $\tilde{m} \sim \{\tilde{m}_1, \tilde{m}_7, \tilde{m}_{27} \}$ for the parameters in eq.~\eqref{alpha-q}, in this model we can again get the correct Higgs mass which is not sensitive to colored top partners and satisfy minimal tuning for masses
\begin{equation}\label{prediction-light-quarks}
\begin{split}
&\tilde{m} \sim (4-6) f \quad \mathrm{bottom},\\
&\tilde{m} \sim (7-9) f \quad \mathrm{charm},
\end{split}
\end{equation}
using the values of $y_b$ and $y_c$ at the TeV scale.  Above these reference values the tuning grows as $f^2/v^2 (m_\Psi/\tilde{m})^3$ as in eq.~\eqref{tuning-lightquarks}.   These values are estimated such that minimal tuning is satisfied.

The simplest interpretation is that the overall scale $\tilde{m}$ is the same for both the lighter quarks and the top, however we could make the further assumption that $\tilde{m}$ be unrelated to the mass scale of the top partners.  This means that only the light quark partners need to satisfy eq.~\ref{prediction-light-quarks}, and the mass scale of the top partners is ``unconstrained.''  This seems particularly motivated in the case of the charm, as one can then realize a U(2)$^3$ flavor symmetry~\cite{Barbieri:2012uh,Matsedonskyi:2014iha}.  Contrary to Model A, here the overall mass scale of the vector resonances is unconstrained by naturalness, since it contributes to the potential only logarithmically and at subleading order $\order(g^4 f^4)$.

\section{Phenomenology}\label{sec:pheno}

In this section we briefly discuss the phenomenology associated with this class of models.  Despite different numerics in the specific realizations we have presented, the generic prediction is a scale for the composite resonances which is parameterically larger than the Goldstone scale $f$.  This suggests that these models can remain hidden after the second run of the LHC while still being minimally tuned.  In this way, they provide an example of a natural theory that has clear signals at a future collider, while being difficult to discover at the LHC.  There several ways to look for these models.

\paragraph{Indirect searches:}

In the absence of constraints from direct searches of resonances, important complementary information will come from precision measurements.  As the scale $f$ is lower than the resonance masses, $f$ can be probed in several ways.

\begin{itemize}

\item Higgs decay to visible particles.

In this model there is a universal rescaling of all the Higgs couplings eq.~\eqref{couplings}~\cite{Burdman:2014zta} of $\sqrt{1-v^2/f^2}$. Projections available for LHC14 with 300 fb$^{-1}$~\cite{ATLAS-Collaboration:2012jwa,CMS-projections} suggest that $f$ will be probed up to $700 \div 800$ GeV.  We have used the target value of $f=800$ GeV throughout this work.

\item Higgs decay to mirror particles.

The interplay between the Higgs boson and the mirror sector is also interesting, and it can be studied through the Higgs to mirror couplings in eq.~\eqref{couplings}, which, differently from the couplings to SM fields, are more model dependent. Upon EWSB the Higgs couples to mirror particles, and depending on their masses this could result in a contribution to the invisible/undetected Higgs width.  Recall that the expected size of the masses of mirror particles 
\begin{equation}
m_{W'} \sim g' f, \quad\quad m_{f'} \sim y_f' f, \nn
\end{equation}
where $g'$ and $y_f'$ are mirror gauge and Yukawa couplings, given that we allow for a generic $Z_2$-breaking.  The presence of this decay channel, assuming no new physics in the loop, induces a universal rescaling factor to all Higgs signal strengths,
\begin{equation}
\mu=(1-v^2/f^2)(1-\mathrm{BR_{inv}}).
\end{equation}
All mirror particles below $\sim m_h/2$ can contribute to $\mathrm{BR_{inv}}$. It is important to avoid contribution from light mirror vectors, as they contribute with a width,
\begin{equation}
\Gamma(h\to V'V') \sim \frac{v^4}{f^4} \times \left(1 - \frac{v^2}{f^2}\right) \times \frac{m_h}{8\pi}\left(\frac{m_h}{v}\right)^2, \nn
\end{equation}
which, despite the suppression $v^4/f^4$, can be numerically relevant.  That is the reason why in the analysis of figure~\ref{fig:tuning} we restricted to values of $m_{W'}$ safely large. In the limit of a small $Z_2$-breaking in the gauge sector, the next important channel to look at is the Higgs decay to pairs of mirror bottom quarks.  Barring kinematical factors, the following relation approximatively holds
\begin{equation}
\Gamma(h \to b'\bar{b}') \simeq \frac{y_b^2}{y_b'^2} \frac{v^2}{f^2}\Gamma(h\to b\bar{b})=\frac{y_b^2}{y_b'^2} \frac{v^2}{f^2}\left(1-\frac{v^2}{f^2}\right)\Gamma(h\to b\bar{b})_{\rm SM},
\end{equation}
where the first approximation depends on the different masses of the bottom and the mirror bottom which differ by $y_b/y_b' \times v/f$. For reasonable values of $f$ and in the almost exact $Z_2$ limit, the width can be sizable.  In this limit the $\mathrm{BR_{inv}}$ is reasonably dominated by the decay to a pair of mirror bottoms, and it is of the order $5 \div 10\%$ for $f=800$ GeV (see also \cite{Craig:2015pha}).

\item Electroweak precision tests.

Another indirect constraint -- which is rather model independent -- comes from the IR-logarithms in the $S$ and $T$ parameters of the electroweak precision tests~\cite{Barbieri:2007bh}.  While these contributions can be sizable, it has been shown~\cite{Grojean:2013qca,Azatov:2013ura} that lower values of $f$ can reconciled with data by invoking UV contributions from the composite sector.  However, a value of $f \gtrsim 800$ GeV can be considered as the target precision for CTH models. 
\end{itemize}

\paragraph{Direct searches:}

In order to evaluate the robustness of our setup, it is important to recall the projections for direct searches of composite resonances at the LHC
\begin{itemize}
\item Colored top partners.

  Currently top partners searches look for top partners produced via pair production and at LHC14 with  300 fb$^{-1}$ the mass reach is $m_\Psi \gtrsim 2$ TeV (and $\sim 2.8$ TeV at 3 ab$^{-1}$)~\cite{Matsedonskyi:2014mna}, which is easily satisfied in the models we have presented.  More promising are searches for singly produced top partners, despite the fact that these kind of searches have not yet been performed, individual studies~\cite{Ortiz:2014iza,Matsedonskyi:2014mna} show that they can be relevant at higher masses (see also \cite{Backovic:2014uma,Backovic:2015lfa}).  They are more model dependent, but even in the most favorable parameter space may put a bound of $m_\Psi \gtrsim 3$ TeV at the end of the LHC program.

  Note that the light fermion resonances are allowed in models, because when $Z_2$ is broken in the gauge sector the fermion resonances are largely uncorrelated with tuning.  In this case, however, the twin mechanism becomes largely superfluous.

\item Vector resonances.

The searches for vector resonances (see~\cite{Pappadopulo:2014qza} for a discussion) will probe portions of the natural parameter space of the model with $Z_2$-breaking only in the gauge sector, as shown for example in figure~\ref{fig:tuning} (for $f=800$ GeV), but they will leave all other models unaffected. However, a lower bound $m_\rho \gtrsim 2.5$ TeV is generically required by the EWPT, see e.g. \cite{Contino:2010rs}.

\item The mirror top. 

  The mirror top plays a key role in the generation of the Higgs potential and is expected to lie in the TeV range, with a mass $m_{t'} \sim y_{t'} f$.  They can be pair produced through an off-shell Higgs, but their signature is very dependent on the dynamics of the mirror sector.  Their decay chains could occur entirely in the mirror sector, resulting in missing energy, or partially back into SM particles via an off-shell Higgs, resulting in either soft SM particles or displaced vertices.  An overview of some of the wide-range of possibilities is given in~\cite{Craig:2015pha}.  Ref.~\cite{Craig:2014lda} presents a search for SM singlets pair produced through an off-shell Higgs and finds the reach to be $\sim 200$ GeV at the LHC and $\sim 300$ GeV at a 100 TeV collider.

\end{itemize}

\section{Conclusions}\label{sec:conclusions}

In this paper, we have discussed the tuning of the electroweak vacuum and the Higgs in the composite Higgs framework augmented with the twin Higgs mechanism.  In the simplest realization the Higgs is a pNGB of SO(8)/SO(7) with a potential generated by its couplings to elementary quarks and gauge fields.  The twin Higgs mechanism makes the generated potential symmetric under a $Z_2$ symmetry which forbids terms quadratic in $m_\Psi$, the scale of fermionic resonances.  We emphasized that removing the sensitivity of the potential to $m_\Psi$ crucially depends both on the $Z_2$ symmetry \textit{and} the $t_R$ being a total singlet of the unbroken global symmetry.  With these pieces in place, the scaling $y^4 f^4$ of the Higgs potential is really $y^4 f^4 \simeq y_t^2 m_{t'}^2 f^2$, indicating that the top contribution to the Higgs potential does not require light colored particles.

\begin{figure}[t]
\begin{center}
\hspace{-50px}
\includegraphics[width=.7\textwidth]{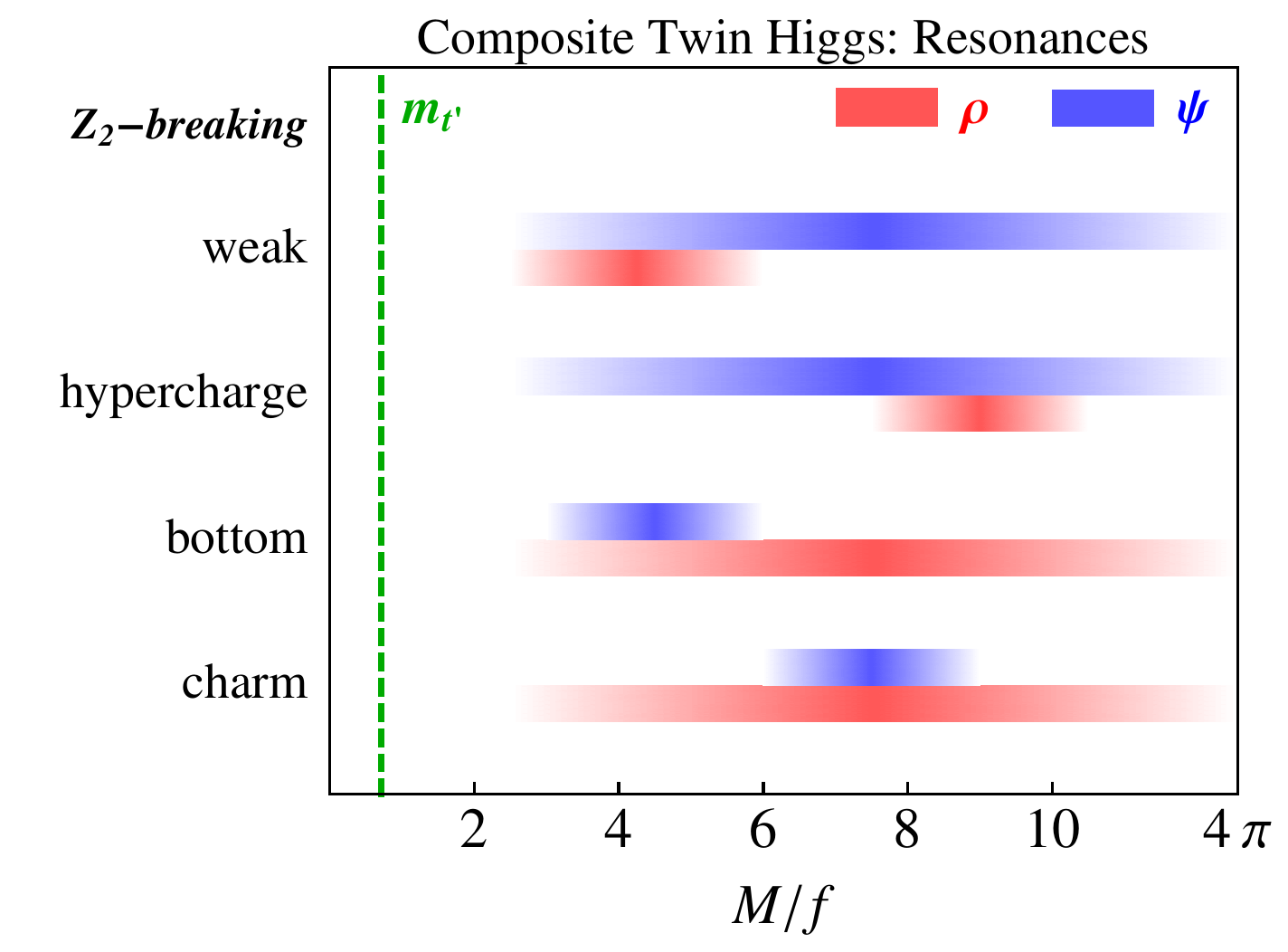}
\caption{Summary of the mass of composite resonances for Composite Twin Higgs for various $Z_2$-breaking mechanisms and minimal tuning. The first case reflects figure~\ref{fig:tuning} and the second \cite{Barbieri:2015lqa} is obtained by a simple rescaling.  They both have an unconstrained fermionic scale with a `predicted' range for the masses of composite vector resonances. The last two models have an unconstrained mass for the vector resonances and a `predicted' range for the fermionic scale (see eq.~\eqref{prediction-light-quarks}).}
\label{fig:resonances}
\end{center}
\end{figure}

As a realistic theory requires $Z_2$ to be broken, we have explored several options.  If the terms that arise due to the $Z_2$-breaking come in at the same order, $y_t^4 f^4$, as the symmetric ones, then the potential really is not sensitive to the colored top partners and minimal tuning, $f^2/v^2$, is satisfied.  We have shown, however, that getting the right $Z_2$-breaking term is not generic in this framework.  For instance, $Z_2$-breaking from the top sector introduces an explicit dependence on $m_\Psi$ and likely spoils the gains from the twin mechanism.  In section~\ref{sec:ewsb} we introduced several mechanisms to break this symmetry and for the more plausible mechanisms of breaking in the gauge sector or in the lighter quarks we made numerical estimates of where the resonances would lie, under minimal tuning.  Figure~\ref{spectrum} provides a cartoon of the predicted spectrum relative to standard composite Higgs and figure~\ref{fig:resonances} summarizes the rough spectrum depending on the $Z_2$-breaking mechanism.

In this work we have identified multiple $Z_2$-breaking methods and explored their individual spectra.  There is no reason, however, why more than one mechanism cannot be at work simultaneously.  We expect that this possibility opens up even more parameter space that may not be populated by the examples in figure~\ref{fig:resonances}.  This would be an interesting scenario which we leave for future work.

We discussed also the phenomenology of these models. If the spectrum as in figure~\ref{fig:resonances} is taken at face value, the phenomenology is mainly controlled by the goldstone scale $f$, the only scale nearby.  Direct searches, especially for the light composite vector resonances, may foray into regions of the parameter space, but in all models presented minimally tuned regions will survive.  On the other hand, precision tests of the Higgs couplings will constrain the models, especially in the final stages of second run of the LHC.  Here the discussion closely resembles the one for a standard composite Higgs, where the improvements on the couplings will probe $f$ up to 800 GeV at LHC14 with 300 fb$^{-1}$.  However, differently from the standard case, here a complementary opportunity is offered by the phenomenology of the mirror particles in their contribution to the invisible decay width of the Higgs. 

To conclude, given that we have outlined several possible scenarios for realistic Composite Twin Higgs models, it would be nice to fully explore their experimental signatures.

\subsubsection*{Acknowledgements}

The authors would like to thank Davide Greco, Riccardo Rattazzi, Andrea Wulzer, and especially Riccardo Barbieri for useful discussions related to their forthcoming work~\cite{Barbieri:2015lqa}. AT is supported by an Oheme Fellowship and LTW is supported by DOE grant DE-SC0003930.

\subsubsection*{Note added}

Recently, a related paper appeared that discussed the Twin Higgs~\cite{Craig:2015pha}.  Although there are similarities, our approach differs in the global symmetries and in the fact that we use the twin Higgs mechanism within a calculable composite Higgs model.

\appendix
\section{Technicalities of SO(8)/SO(7)}\label{sec:app}

\subsubsection*{Generators}

Defining $t_{ij}^{ab}= \delta^a_i\delta^b_j-\delta^a_j\delta^b_i$, the vector representation of the SO(8) generators is given by
\begin{equation}
  T^{ab}_{ij} = -\frac{i}{\sqrt{2}}t_{ij}^{ab}
  \quad\quad\quad
  \begin{array}{l}
  a = 1,\ldots,8, \\
  b = a+1,\ldots,8,
  \end{array}
\end{equation}
such that there are 28 generators.  The normalization ${\rm Tr}(T^a T^b)=\delta^{ab}$ is used throughout.  We define the embedding of the two SO(4) subgroups in the following way
\begin{equation}
\begin{split}
  (T ^{a}_{ij})_{L,R} &= -\frac{i}{2}\big( \frac{1}{2}\epsilon^{abc}t_{ij}^{bc}\pm t^{a4}_{ij} \big) \quad\quad a=1,2,3, \\
  (T'^{a}_{ij})_{L,R} &= -\frac{i}{2}\big( \frac{1}{2} \epsilon^{abc}t_{ij}^{bc}\pm t^{a8}_{ij} \big) \quad\quad a=5,6,7,
\end{split}
\end{equation}
where the unprimed generators denote the SO(4) containing the standard model and the primed generators denote the mirror SO(4).  The breaking of SO(8) to SO(7) results in 7 broken generators:
\begin{equation}
T^{\hat{a}}_{ij}= -\frac{i}{\sqrt{2}}t^{\hat{a}8}_{ij}, \quad\quad \hat{a}=1,\ldots,7.
\end{equation}
The pion field is obtained via exponentiation $U(\Pi)=\exp i\Pi(x)/f$, where $\Pi(x) = \sqrt{2}h^{\hat{a}}(x) T^{\hat{a}}$.

\subsubsection*{Fermion representations}

Once the embedding of the gauge groups has been fixed we can describe the embedding of quarks into incomplete representations of SO(8).  Below we show the embedding of the visible sector, the mirror quarks will be the same given the mirror exchange.  The $q_L$ must be embedded in the \textbf{8}, while the $u_R$ can be in the \textbf{1}, \textbf{28}, or \textbf{35}
\begin{itemize}

\item $q_L$ in the \textbf{8}:
\begin{equation}
(q_L^{\mathbf{8}})^i = \frac{1}{\sqrt{2}}\left(i b_L, b_L, i t_L, -t_L, 0, 0, 0, 0 \right)^i.
\end{equation}

\item $u_R$ in the \textbf{1} is a total singlet of SO(8).

\item $u_R$ in the \textbf{28}:
\begin{equation}
(u_R^{\mathbf{28}})^{ij}= \left(
\begin{array}{cccccccc}
 0 & \frac{u_R}{2} & 0 & 0 & 0 & 0 & 0 & 0 \\
 -\frac{u_R}{2} & 0 & 0 & 0 & 0 & 0 & 0 & 0 \\
 0 & 0 & 0 & \frac{u_R}{2} & 0 & 0 & 0 & 0 \\
 0 & 0 & -\frac{u_R}{2} & 0 & 0 & 0 & 0 & 0 \\
 0 & 0 & 0 & 0 & 0 & 0 & 0 & 0 \\
 0 & 0 & 0 & 0 & 0 & 0 & 0 & 0 \\
 0 & 0 & 0 & 0 & 0 & 0 & 0 & 0 \\
 0 & 0 & 0 & 0 & 0 & 0 & 0 & 0 \\
\end{array}
\right)^{ij}.
\end{equation}

\item $u_R$ can be embedded in the \textbf{35}:
\begin{equation}
  (u_R^{\mathbf{35}})^{ij} = \frac{u_R}{\sqrt{8}}\mathrm{diag}(1,1,1,1,-1,-1,-1,-1)^{ij}.
\end{equation}
From this expression one can see that it is a singlet separately for each SO(4) subgroup.

\end{itemize}

\subsubsection*{Functional form of the potential} 

In order to understand the origin of Table~\ref{tab:functions}, one should construct all possible invariants composed of goldstones $\Sigma$, and spurions $y_{L,R}$~\cite{Mrazek:2011iu}.  In the effective lagrangian, after the composite sector has been integrated out, one finds a number of terms which can be classified according to whether they originate from a correction to a kinetic term or a mass term.  We restrict the discussion to the visible sector, the mirror terms can be found using the usual substitution of $s_h \to c_h$.  The results are shown in Table~\ref{tab:potform}.

\begin{table}[h]
\begin{center}
\begin{tabular}{| c c c c |} \hline
  Origin & Fields & Order & Forms \\ \hline \hline
  kinetic term & $q_L^\mathbf{8}$  & $\order(y_L^2,y_L^4)$ & $y_L^2 s_h^2$, $y_L^4 s_h^4$  \\
  kinetic term  & $u_R^\mathbf{1}$  & $\order(y_R^2,y_R^4)$ & -- \\
  kinetic term & $u_R^\mathbf{28}$ & $\order(y_R^2,y_R^4)$ & $y_R^2 s_h^2$, $y_R^4 s_h^4$  \\
  kinetic term & $u_R^\mathbf{35}$ & $\order(y_R^2,y_R^4)$ & $y_R^2 c_{2h}^2$, $y_R^4 c_{2h}^4$  \\
  mass term & $q_L^\mathbf{8}$, $u_R^\mathbf{1}$  & $\order(y_L^2)$ & $y_L^2 s_h^2$  \\
  mass term & $q_L^\mathbf{8}$, $u_R^\mathbf{28}$ & $\order(y_L^2 y_R^2)$ & $y_L^2 y_R^2 s_h^2$  \\
  mass term & $q_L^\mathbf{8}$, $u_R^\mathbf{35}$ & $\order(y_L^2 y_R^2)$ & $y_L^2 y_R^2 s_h^2$, $y_L^2 y_R^2 c_{2h}^2$  \\ \hline
\end{tabular}
\caption{\label{tab:potform} Structures that can appear in the Higgs potential (from the visible sector).}
\end{center}
\end{table}
From Table~\ref{tab:potform} one can see that the $Z_2$-symmetric potential for the above cases has the following scaling with the elementary-composite mixings $y_L$ and $y_R$,
\begin{equation}\label{various-potentials}
\begin{split}
V_{q_L^{\mathbf{8}}+u_R^{\mathbf{1}}} &\sim \order(y_L^4) s_h^2 c_h^2,\\
V_{q_L^{\mathbf{8}}+u_R^{\mathbf{28}}} &\sim \order(y_L^4,y_R^4) s_h^2 c_h^2,\\
V_{q_L^{\mathbf{8}}+u_R^{\mathbf{35}}} &\sim \order(y_R^2,y_L^4,y_L^2y_R^2)s_h^2 c_h^2.
\end{split}
\end{equation}

\pagestyle{plain}
\bibliographystyle{jhep}
\small
\bibliography{ref}

\providecommand{\href}[2]{#2}\begingroup\raggedright\begin{thebibliography}{10}

\bibitem{Papucci:2011wy}
M.~Papucci, J.~T. Ruderman, and A.~Weiler, {\it {Natural SUSY Endures}},  {\em
  JHEP} {\bf 1209} (2012) 035, [\href{http://arxiv.org/abs/1110.6926}{{\tt
  arXiv:1110.6926}}].

\bibitem{Hall:2011aa}
L.~J. Hall, D.~Pinner, and J.~T. Ruderman, {\it {A Natural SUSY Higgs Near 126
  GeV}},  {\em JHEP} {\bf 1204} (2012) 131,
  [\href{http://arxiv.org/abs/1112.2703}{{\tt arXiv:1112.2703}}].

\bibitem{Dugan:1984hq}
M.~J. Dugan, H.~Georgi, and D.~B. Kaplan, {\it {Anatomy of a Composite Higgs
  Model}},  {\em Nucl.Phys.} {\bf B254} (1985) 299.

\bibitem{Kaplan:1991dc}
D.~B. Kaplan, {\it {Flavor at SSC energies: A New mechanism for dynamically
  generated fermion masses}},  {\em Nucl.Phys.} {\bf B365} (1991) 259--278.

\bibitem{Contino:2006nn}
R.~Contino, T.~Kramer, M.~Son, and R.~Sundrum, {\it {Warped/composite
  phenomenology simplified}},  {\em JHEP} {\bf 0705} (2007) 074,
  [\href{http://arxiv.org/abs/hep-ph/0612180}{{\tt hep-ph/0612180}}].

\bibitem{Agashe:2004rs}
K.~Agashe, R.~Contino, and A.~Pomarol, {\it {The Minimal composite Higgs
  model}},  {\em Nucl.Phys.} {\bf B719} (2005) 165--187,
  [\href{http://arxiv.org/abs/hep-ph/0412089}{{\tt hep-ph/0412089}}].

\bibitem{Contino:2006qr}
R.~Contino, L.~Da~Rold, and A.~Pomarol, {\it {Light custodians in natural
  composite Higgs models}},  {\em Phys.Rev.} {\bf D75} (2007) 055014,
  [\href{http://arxiv.org/abs/hep-ph/0612048}{{\tt hep-ph/0612048}}].

\bibitem{Contino:2010rs}
R.~Contino, {\it {The Higgs as a Composite Nambu-Goldstone Boson}},
  \href{http://arxiv.org/abs/1005.4269}{{\tt arXiv:1005.4269}}.

\bibitem{Matsedonskyi:2012ym}
O.~Matsedonskyi, G.~Panico, and A.~Wulzer, {\it {Light Top Partners for a Light
  Composite Higgs}},  {\em JHEP} {\bf 1301} (2013) 164,
  [\href{http://arxiv.org/abs/1204.6333}{{\tt arXiv:1204.6333}}].

\bibitem{Redi:2012ha}
M.~Redi and A.~Tesi, {\it {Implications of a Light Higgs in Composite Models}},
   {\em JHEP} {\bf 1210} (2012) 166,
  [\href{http://arxiv.org/abs/1205.0232}{{\tt arXiv:1205.0232}}].

\bibitem{Marzocca:2012zn}
D.~Marzocca, M.~Serone, and J.~Shu, {\it {General Composite Higgs Models}},
  {\em JHEP} {\bf 1208} (2012) 013, [\href{http://arxiv.org/abs/1205.0770}{{\tt
  arXiv:1205.0770}}].

\bibitem{Pomarol:2012qf}
A.~Pomarol and F.~Riva, {\it {The Composite Higgs and Light Resonance
  Connection}},  {\em JHEP} {\bf 1208} (2012) 135,
  [\href{http://arxiv.org/abs/1205.6434}{{\tt arXiv:1205.6434}}].

\bibitem{Panico:2012uw}
G.~Panico, M.~Redi, A.~Tesi, and A.~Wulzer, {\it {On the Tuning and the Mass of
  the Composite Higgs}},  {\em JHEP} {\bf 1303} (2013) 051,
  [\href{http://arxiv.org/abs/1210.7114}{{\tt arXiv:1210.7114}}].

\bibitem{Schmaltz:2005ky}
M.~Schmaltz and D.~Tucker-Smith, {\it {Little Higgs review}},  {\em
  Ann.Rev.Nucl.Part.Sci.} {\bf 55} (2005) 229--270,
  [\href{http://arxiv.org/abs/hep-ph/0502182}{{\tt hep-ph/0502182}}].

\bibitem{Bellazzini:2014yua}
B.~Bellazzini, C.~Csáki, and J.~Serra, {\it {Composite Higgses}},  {\em
  Eur.Phys.J.} {\bf C74} (2014), no.~5 2766,
  [\href{http://arxiv.org/abs/1401.2457}{{\tt arXiv:1401.2457}}].

\bibitem{Chacko:2005pe}
Z.~Chacko, H.-S. Goh, and R.~Harnik, {\it {The Twin Higgs: Natural electroweak
  breaking from mirror symmetry}},  {\em Phys.Rev.Lett.} {\bf 96} (2006)
  231802, [\href{http://arxiv.org/abs/hep-ph/0506256}{{\tt hep-ph/0506256}}].

\bibitem{Chacko:2005un}
Z.~Chacko, H.-S. Goh, and R.~Harnik, {\it {A Twin Higgs model from left-right
  symmetry}},  {\em JHEP} {\bf 0601} (2006) 108,
  [\href{http://arxiv.org/abs/hep-ph/0512088}{{\tt hep-ph/0512088}}].

\bibitem{Barbieri:2005ri}
R.~Barbieri, T.~Gregoire, and L.~J. Hall, {\it {Mirror world at the large
  hadron collider}},  \href{http://arxiv.org/abs/hep-ph/0509242}{{\tt
  hep-ph/0509242}}.

\bibitem{Falkowski:2006qq}
A.~Falkowski, S.~Pokorski, and M.~Schmaltz, {\it {Twin SUSY}},  {\em Phys.Rev.}
  {\bf D74} (2006) 035003, [\href{http://arxiv.org/abs/hep-ph/0604066}{{\tt
  hep-ph/0604066}}].

\bibitem{Chang:2006ra}
S.~Chang, L.~J. Hall, and N.~Weiner, {\it {A Supersymmetric twin Higgs}},  {\em
  Phys.Rev.} {\bf D75} (2007) 035009,
  [\href{http://arxiv.org/abs/hep-ph/0604076}{{\tt hep-ph/0604076}}].

\bibitem{Craig:2013fga}
N.~Craig and K.~Howe, {\it {Doubling down on naturalness with a supersymmetric
  twin Higgs}},  {\em JHEP} {\bf 1403} (2014) 140,
  [\href{http://arxiv.org/abs/1312.1341}{{\tt arXiv:1312.1341}}].

\bibitem{Craig:2014aea}
N.~Craig, S.~Knapen, and P.~Longhi, {\it {Neutral Naturalness from the Orbifold
  Higgs}},  \href{http://arxiv.org/abs/1410.6808}{{\tt arXiv:1410.6808}}.

\bibitem{Craig:2014roa}
N.~Craig, S.~Knapen, and P.~Longhi, {\it {The Orbifold Higgs}},
  \href{http://arxiv.org/abs/1411.7393}{{\tt arXiv:1411.7393}}.

\bibitem{Burdman:2014zta}
G.~Burdman, Z.~Chacko, R.~Harnik, L.~de~Lima, and C.~B. Verhaaren, {\it
  {Colorless Top Partners, a 125 GeV Higgs, and the Limits on Naturalness}},
  \href{http://arxiv.org/abs/1411.3310}{{\tt arXiv:1411.3310}}.

\bibitem{Geller:2014kta}
M.~Geller and O.~Telem, {\it {A Holographic Twin Higgs Model}},
  \href{http://arxiv.org/abs/1411.2974}{{\tt arXiv:1411.2974}}.

\bibitem{Barbieri:2015lqa}
R.~Barbieri, D.~Greco, R.~Rattazzi, and A.~Wulzer, {\it {The Composite Twin
  Higgs scenario}},  \href{http://arxiv.org/abs/1501.07803}{{\tt
  arXiv:1501.07803}}.

\bibitem{Batra:2008jy}
P.~Batra and Z.~Chacko, {\it {A Composite Twin Higgs Model}},  {\em Phys.Rev.}
  {\bf D79} (2009) 095012, [\href{http://arxiv.org/abs/0811.0394}{{\tt
  arXiv:0811.0394}}].

\bibitem{DeSimone:2012fs}
A.~De~Simone, O.~Matsedonskyi, R.~Rattazzi, and A.~Wulzer, {\it {A First Top
  Partner Hunter's Guide}},  {\em JHEP} {\bf 1304} (2013) 004,
  [\href{http://arxiv.org/abs/1211.5663}{{\tt arXiv:1211.5663}}].

\bibitem{Craig:2015pha}
N.~Craig, A.~Katz, M.~Strassler, and R.~Sundrum, {\it {Naturalness in the Dark
  at the LHC}},  \href{http://arxiv.org/abs/1501.05310}{{\tt
  arXiv:1501.05310}}.

\bibitem{Coleman:1973jx}
S.~R. Coleman and E.~J. Weinberg, {\it {Radiative Corrections as the Origin of
  Spontaneous Symmetry Breaking}},  {\em Phys.Rev.} {\bf D7} (1973) 1888--1910.

\bibitem{Giudice:2007fh}
G.~Giudice, C.~Grojean, A.~Pomarol, and R.~Rattazzi, {\it {The
  Strongly-Interacting Light Higgs}},  {\em JHEP} {\bf 0706} (2007) 045,
  [\href{http://arxiv.org/abs/hep-ph/0703164}{{\tt hep-ph/0703164}}].

\bibitem{Mrazek:2011iu}
J.~Mrazek, A.~Pomarol, R.~Rattazzi, M.~Redi, J.~Serra, and A.~Wulzer, {\it {The
  Other Natural Two Higgs Doublet Model}},  {\em Nucl.Phys.} {\bf B853} (2011)
  1--48, [\href{http://arxiv.org/abs/1105.5403}{{\tt arXiv:1105.5403}}].

\bibitem{Carmona:2014iwa}
A.~Carmona and F.~Goertz, {\it {A naturally light Higgs without light Top
  Partners}},  \href{http://arxiv.org/abs/1410.8555}{{\tt arXiv:1410.8555}}.

\bibitem{Panico:2011pw}
G.~Panico and A.~Wulzer, {\it {The Discrete Composite Higgs Model}},  {\em
  JHEP} {\bf 1109} (2011) 135, [\href{http://arxiv.org/abs/1106.2719}{{\tt
  arXiv:1106.2719}}].

\bibitem{DeCurtis:2011yx}
S.~De~Curtis, M.~Redi, and A.~Tesi, {\it {The 4D Composite Higgs}},  {\em JHEP}
  {\bf 1204} (2012) 042, [\href{http://arxiv.org/abs/1110.1613}{{\tt
  arXiv:1110.1613}}].

\bibitem{Barbieri:1987fn}
R.~Barbieri and G.~Giudice, {\it {Upper Bounds on Supersymmetric Particle
  Masses}},  {\em Nucl.Phys.} {\bf B306} (1988) 63.

\bibitem{ATLAS-Collaboration:2012jwa}
T.~ATLAS-Collaboration, {\it {Physics at a High-Luminosity LHC with ATLAS}},
  Tech. Rep. ATL-PHYS-PUB-2012-001, CERN, Geneva, Aug, 2012.

\bibitem{CMS-projections}
{\bf CMS Collaboration} Collaboration, {\it {CMS at the High-Energy Frontier.
  Contribution to the Update of the European Strategy for Particle Physics}},
  Tech. Rep. CMS-NOTE-2012-006. CERN-CMS-NOTE-2012-006, CERN, Geneva, Oct,
  2012.

\bibitem{Contino:2013gna}
R.~Contino, C.~Grojean, D.~Pappadopulo, R.~Rattazzi, and A.~Thamm, {\it {Strong
  Higgs Interactions at a Linear Collider}},  {\em JHEP} {\bf 1402} (2014) 006,
  [\href{http://arxiv.org/abs/1309.7038}{{\tt arXiv:1309.7038}}].

\bibitem{Barbieri:2012uh}
R.~Barbieri, D.~Buttazzo, F.~Sala, and D.~M. Straub, {\it {Flavour physics from
  an approximate U(2)3 symmetry}},  {\em JHEP} {\bf 1207} (2012) 181,
  [\href{http://arxiv.org/abs/1203.4218}{{\tt arXiv:1203.4218}}].

\bibitem{Matsedonskyi:2014iha}
O.~Matsedonskyi, {\it {On Flavour and Naturalness of Composite Higgs Models}},
  \href{http://arxiv.org/abs/1411.4638}{{\tt arXiv:1411.4638}}.

\bibitem{Barbieri:2007bh}
R.~Barbieri, B.~Bellazzini, V.~S. Rychkov, and A.~Varagnolo, {\it {The Higgs
  boson from an extended symmetry}},  {\em Phys.Rev.} {\bf D76} (2007) 115008,
  [\href{http://arxiv.org/abs/0706.0432}{{\tt arXiv:0706.0432}}].

\bibitem{Grojean:2013qca}
C.~Grojean, O.~Matsedonskyi, and G.~Panico, {\it {Light top partners and
  precision physics}},  {\em JHEP} {\bf 1310} (2013) 160,
  [\href{http://arxiv.org/abs/1306.4655}{{\tt arXiv:1306.4655}}].

\bibitem{Azatov:2013ura}
A.~Azatov, R.~Contino, A.~Di~Iura, and J.~Galloway, {\it {New Prospects for
  Higgs Compositeness in $h \to Z\gamma$}},  {\em Phys.Rev.} {\bf D88} (2013),
  no.~7 075019, [\href{http://arxiv.org/abs/1308.2676}{{\tt arXiv:1308.2676}}].

\bibitem{Matsedonskyi:2014mna}
O.~Matsedonskyi, G.~Panico, and A.~Wulzer, {\it {On the Interpretation of Top
  Partners Searches}},  {\em JHEP} {\bf 1412} (2014) 097,
  [\href{http://arxiv.org/abs/1409.0100}{{\tt arXiv:1409.0100}}].

\bibitem{Ortiz:2014iza}
N.~G. Ortiz, J.~Ferrando, D.~Kar, and M.~Spannowsky, {\it {Reconstructing
  singly produced top partners in decays to Wb}},  {\em Phys.Rev.} {\bf D90}
  (2014), no.~7 075009, [\href{http://arxiv.org/abs/1403.7490}{{\tt
  arXiv:1403.7490}}].

\bibitem{Backovic:2014uma}
M.~Backović, G.~Perez, T.~Flacke, and S.~J. Lee, {\it {LHC Top Partner
  Searches Beyond the 2 TeV Mass Region}},
  \href{http://arxiv.org/abs/1409.0409}{{\tt arXiv:1409.0409}}.

\bibitem{Backovic:2015lfa}
M.~Backović, T.~Flacke, J.~H. Kim, and S.~J. Lee, {\it {Discovering Heavy New
  Physics in Boosted $Z$ Channels: $Z \rightarrow l ^+l^- $ vs. $ Z \rightarrow
  \nu \bar{\nu}$}},  \href{http://arxiv.org/abs/1501.07456}{{\tt
  arXiv:1501.07456}}.

\bibitem{Pappadopulo:2014qza}
D.~Pappadopulo, A.~Thamm, R.~Torre, and A.~Wulzer, {\it {Heavy Vector Triplets:
  Bridging Theory and Data}},  {\em JHEP} {\bf 1409} (2014) 060,
  [\href{http://arxiv.org/abs/1402.4431}{{\tt arXiv:1402.4431}}].

\bibitem{Craig:2014lda}
N.~Craig, H.~K. Lou, M.~McCullough, and A.~Thalapillil, {\it {The Higgs Portal
  Above Threshold}},  \href{http://arxiv.org/abs/1412.0258}{{\tt
  arXiv:1412.0258}}.

\end{thebibliography}\endgroup
\end{document}